\documentclass[12pt]{article}
\usepackage{graphicx}
\usepackage{url}

\newcommand\pubnumber{SLAC--PUB--17154}
\newcommand\pubdate{\hfill September, 2017}

\def\SLAC{SLAC,
    Stanford University, Menlo Park, California 94025 USA}
\def\doeack{\footnote{Work supported by the US Department of Energy,
                     contract DE--AC02--76SF00515.}}

\def\Title#1{\begin{center} {\Large #1 } \end{center}}
\def\Author#1{\begin{center}{ \sc #1} \end{center}}
\def\Address#1{\begin{center}{ \it #1} \end{center}}

\def\submit#1{\begin{center}Submitted to {\sl #1} \end{center}}
\newcommand\pubblock{\rightline{\begin{tabular}{l} \pubnumber\\
         \pubdate \end{tabular}}}
\newenvironment{Abstract}{\begin{quotation} \begin{center}
                       ABSTRACT
     \end{center}\bigskip  }{\end{quotation}}

\def\submit#1{\begin{center}Submitted to {\sl #1} \end{center}}
\def\Acknowledgements{\bigskip  \bigskip \begin{center} \begin{large}
             \bf ACKNOWLEDGEMENTS \end{large}\end{center}}
\def\beq{\begin{equation}}
\def\eeq#1{\label{#1}\end{equation}}
\def\eeqn{\end{equation}}

\newenvironment{Eqnarray}%
   {\arraycolsep 0.14em\begin{eqnarray}}{\end{eqnarray}}
\def\beqa{\begin{Eqnarray}}
\def\eeqa#1{\label{#1}\end{Eqnarray}}
\def\eeqan{\end{Eqnarray}}
\def\CR{\nonumber \\ }

\def\leqn#1{(\ref{#1})}

\let\bar=\overbar

\def\VEV#1{\left\langle{ #1} \right\rangle}

\def\lsim{\mathrel{\raise.3ex\hbox{$<$\kern-.75em\lower1ex\hbox{$\sim$}}}}
\def\gsim{\mathrel{\raise.3ex\hbox{$>$\kern-.75em\lower1ex\hbox{$\sim$}}}}

\def\tr{{\mbox{\rm tr}}}
\def\half{\frac{1}{2}}

\def\del{\partial}
\def\Dslash{\not{\hbox{\kern-4pt $D$}}}
\def\dslash{\not{\hbox{\kern-2pt $\del$}}}

\def\msb{{\bar{\scriptsize M \kern -1pt S}}}

\def\drb{{\bar{\scriptsize D \kern -1pt R}}}

\def\eps{\epsilon}

\makeatletter
\def\section{\@startsection{section}{0}{\z@}{5.5ex plus .5ex minus
 1.5ex}{2.3ex plus .2ex}{\large\bf}}
\def\subsection{\@startsection{subsection}{1}{\z@}{3.5ex plus .5ex minus
 1.5ex}{1.3ex plus .2ex}{\normalsize\bf}}
\def\subsubsection{\@startsection{subsubsection}{2}{\z@}{-3.5ex plus
-1ex minus  -.2ex}{2.3ex plus .2ex}{\normalsize\sl}}

\renewcommand{\@makecaption}[2]{%
   \vskip 10pt
   \setbox\@tempboxa\hbox{\small #1: #2}
   \ifdim \wd\@tempboxa >\hsize     % IF longer than one line:
       \small #1: #2\par          %   THEN set as ordinary paragraph.
     \else                        %   ELSE  center.
       \hbox to\hsize{\hfil\box\@tempboxa\hfil}
   \fi}

 \def\citenum#1{{\def\@cite##1##2{##1}\cite{#1}}}
\newcount\@tempcntc
\def\@citex[#1]#2{\if@filesw\immediate\write\@auxout{\string\citation{#2}}\fi
  \@tempcnta\z@\@tempcntb\m@ne\def\@citea{}\@cite{\@for\@citeb:=#2\do
    {\@ifundefined
       {b@\@citeb}{\@citeo\@tempcntb\m@ne\@citea\def\@citea{,}{\bf ?}\@warning
       {Citation `\@citeb' on page \thepage \space undefined}}%
    {\setbox\z@\hbox{\global\@tempcntc0\csname b@\@citeb\endcsname\relax}%
     \ifnum\@tempcntc=\z@ \@citeo\@tempcntb\m@ne
       \@citea\def\@citea{,}\hbox{\csname b@\@citeb\endcsname}%
     \else
      \advance\@tempcntb\@ne
      \ifnum\@tempcntb=\@tempcntc
      \else\advance\@tempcntb\m@ne\@citeo
      \@tempcnta\@tempcntc\@tempcntb\@tempcntc\fi\fi}}\@citeo}{#1}}
\def\@citeo{\ifnum\@tempcnta>\@tempcntb\else\@citea\def\@citea{,}%
  \ifnum\@tempcnta=\@tempcntb\the\@tempcnta\else
  {\advance\@tempcnta\@ne\ifnum\@tempcnta=\@tempcntb \else\def\@citea{--}\fi
    \advance\@tempcnta\m@ne\the\@tempcnta\@citea\the\@tempcntb}\fi\fi}
\makeatother

\def\A{{\cal A}}
\def\J{{\cal J}} 
\def\G{{\cal G}}
\def\AA{{\bf A}}
\def\GG{{\bf G}_0}
\def\CC{{\bf C}}
\begin{document}
\begin{titlepage}
\pubblock

\vfill
\Title{Competing Forces in 5-Dimensional Fermion Condensation}
\vfill
\Author{Jongmin Yoon and Michael E. Peskin\doeack}
\Address{\SLAC}
\vfill
\begin{Abstract}
We study  fermion condensation in the Randall-Sundrum background as a
setting for composite Higgs models.  We formalize the computation of
the  Coleman-Weinberg potential and present a simple, general formula.  Using
this tool, we study the competition of fermion multiplets with
different boundary conditions, to find conditions for creating a little
hierarchy with the  Higgs field expectation value much smaller than
the intrinsic Randall-Sundrum mass scale.
 \end{Abstract}
\vfill
\submit{Physical Review D}
\vfill

\newpage
\tableofcontents
\end{titlepage}

\def\thefootnote{\fnsymbol{footnote}}
\setcounter{footnote}{0}

\section{Introduction}

One of the most important open questions in particle physics is to
find an explanation for the spontaneous breaking of the weak
interaction symmetry $SU(2)\times U(1)$.   Ideally, we would like to 
calculate the potential associated with the Higgs boson in terms of a
more fundamental set of parameters.   It is well appreciated that this
is not possible within the Standard Model of particle physics.   This idea
then
motivates models of physics that extend the Standard Model.   

In this
paper, we will study the generation of the Higgs potential in
5-dimensional field theory models.  In these models, the Higgs boson
appears as the fifth component of a gauge field. 
 It has been understood for a long
time that fermions in such models can spontaneously acquire mass, 
driving a breaking of  the gauge symmetry~\cite{Hosotani,Toms}.  
  The fifth dimension can be flat, but
here we will study models in 5-dimensional anti-de Sitter space with
boundaries, as in
the model of Randall and Sundrum~\cite{RS}.  Such a model can be viewed as a
dual description of a strongly-coupled field theory in four
dimensions~\cite{RAP}.  Indeed, the study of these five-dimensional models
potentially gives
a simplified but calculable approach to composite Higgs 
models with strong coupling.

In the original Randall-Sundrum model, a fundamental Higgs field was introduced
as a scalar field living on the  4-dimensional subspace or
brane at the infrared boundary.  However, by introducing the Higgs
field as a fundamental scalar field, this approach gives up any chance
to compute the Higgs potential from deeper principles. 
In this paper, we will consider the
Higgs field to arise as the fifth component of a gauge field in the
5-dimensional bulk, an approach 
called {\it gauge-Higgs unification}~\cite{gaugeHiggsone,gaugeHiggstwo}.
The Higgs potential will be generated dynamically, by integrating out
massive fermion and gauge boson states.
We will nevertheless use the abbreviation RS
to denote this class of models.

RS models of the Higgs sector were studied intensely
about ten years ago, by Agashe, Contino, and  Pomarol~\cite{ACP} and
many others.
However,
many issues were not resolved.   Chief among these is the
understanding of  the various 
hierachies of scales required in these models.   RS models with
dynamical symmetry breaking generated by fermions have three distinct
hierarchies that need to be established.   First, the intrinsically
five-dimensional
or Kaluza-Klein states must be much heavier than Standard Model
particles, including the top quark.  Second, Higgs field vacuum
expectation value must be small compared to its natural scale in the 
five-dimensional theory.  Third, the mass generation for light quarks
and leptons due to the composite Higgs must not generate too large anomalous
values for flavor observables.   We will refer to these requirements
as the {\it KK, $v/f$,} and {\it flavor hierarchies}, respectively.   

In this paper, we will discuss the formalism for symmetry breaking in
RS models of gauge-Higgs unification.   Our goal is to
 present strategies for creating KK and $v/f$
hierarchies.   The KK hierarchy is easier to address.  To create such
a hierarchy, we need to build the Higgs potential from several
different components that naturally have different mass scales.  We
will exhibit some features of fermion condensation in RS models that 
lead to models with this property.

The $v/f$ hierarchy is more difficult to generate.  The Higgs field of
an RS model appears as a field of a nonlinear sigma model,
whose characteristic scale we call $f$.   To obtain a Higgs vacuum
expectation value $v$ much smaller than $f$, we must be near a
second-order phase transition in the phase diagram of the model.  We
will present strategies for obtaining such phase transitions.  
Still,  it will always turn out that a $v/f$ hierarchy requires
a fine-tuning in the model. 

This study will give us ingredients that we can use to construct
realistic theories of strong interactions leading to a composite Higgs
boson.   We will present a model that uses these strategies in a
following
 paper~\cite{YPtwo}.

Our concept for an RS model as a dual to a 4-dimensional strongly
coupled theory of composite Higgs bosons leads to some choices that
are different from those that are conventional in the literature.  We
consider the RS dynamics as modelling an approximately conformal
strong interaction theory that exists at energies above 1 TeV, with an
ultraviolet cutoff at about 100 TeV. These scales will provide the
boundaries of the warped RS geometry, called $z_R$ and $z_0$,
respectively, 
in this paper.   The top quark will play a key
role in this theory in breaking electroweak symmetry, but the other
quarks and leptons will have only weak coupling to the new
dynamics.  We will connect the light quarks and leptons to the Higgs
sector through boundary conditions at 100 TeV.  In this, we view our
construction as a dual of a sort of an extended technicolor (ETC)
theory~\cite{ELETC,DSETC}.   ETC is a scheme that is attractive in principle but has many
problems in practice.   It has proven difficult not only to solve the
problems of ETC but even to find a phenomenological treatment in which
its problems can be swept under the rug.   We hope that RS models will 
at least provide a sufficiently shaggy rug that we can make progress
with  this idea.

The outline of this paper will then be as follows:  In Section 2-4, we
present some basic formalism for computation of the Higgs
potential in RS models, including a simple, general formula for the
computation of the Coleman-Weinberg effective potential~\cite{CW}.
   In Section 5, we review the results of 
Contino, Nomura, and Pomarol~\cite{CNP}
 on symmetry-breaking with one fermion
multiplet, which provide a starting point for our constructions.
  In Section 6,  we explore the idea of competition between fermion
  multiplets with different boundary conditions to create models where 
$v/f \ll 1$.  In Section 7, we present a model containing elements with
intrinsically different scales that can lead to relaxed fine-tuning.
Section 8 gives a summary and some perspective.

\section{Coleman-Weinberg potential in RS models}

In this section, we review the formalism for computing the Higgs
potential in RS models.    For the purpose of this paper, we take a
rather narrow definition: An RS model here will be  a model of gauge
and fermion fields living in the interior of a slice
5-dimensional anti-de Sitter space
\beq
      ds^2 =  {1\over (kz)^2} [ dx^m dx_m - dz^2 ] 
\eeq{metric}
 with nontrivial boundary conditions at $z = z_0$ and $z = z_R$, with
$z_0 < z_R$.    Then $z_0$ gives the position of the ``UV brane'' and
$z_R$ gives the position of the ``IR brane''.   In accord with the
philosophy explained in the Introduction, 
we choose very simple boundary conditions on the
IR brane and build the complexity of the theory using the boundary
conditions on the UV brane.  
Using the perhaps more physical metric
\beq
     ds^2 = e^{-2k x^5} dx^m dx_m  -   (dx^5)^2
\eeq{RSmetric}
we take the  size of the interval in $x^5$ to be $\pi R$.  Then
\beq
        z_0 = {1/k}  \qquad       z_R =    e^{\pi k R}/k   \ . 
\eeq{zzeroRdef}

Because this paper focuses on the
properties of the one-loop potential, we will quote formulae for the
Green's functions of fields in RS in Euclidean space.   Similar formulae apply in Minkowski
space.

In the interior or bulk 5-dimensional region, we will have
spin-$\half$ and spin-1 fields.   The 4-dimensional Higgs field will
appear
as the 5th component of a gauge field in 5~dimensions.  In this paper,
we will notate gauge fields as $A_M^A$, where $M = 0,1,2,3,5$, with
lower case $m = 0,1,2,3$, and $A$ is the gauge group index.
   Fermion fields are 4-component Dirac
fields, which we will decompose as 
\beq
            \Psi =   \pmatrix{\psi_L \cr \psi_R }    \ ,
\eeq{fermiondec}
where $\psi_L$ transforms as a left-handed Weyl fermion
 and $\psi_R$ transforms as a
right-handed Weyl fermion under
4-dimensional Lorentz transformations.  More details of our formalism
for 5-d fermions
are presented in Appendix A.

Quantum fields in the RS geometry were analyzed soon after the
original RS work~\cite{GW,DHR,Grossman}.
Gherghetta and
Pomarol   showed
that fields of all spin values have simple and parallel behavior in
the RS geometry~\cite{GP}.   For a spin 0 field of mass $m$ satisfying the
Klein-Gordon equation,
the solutions in Euclidean space are given by Bessel functions as 
\beq
     \phi(x) = z^2 [  A  I_{\nu} (pz) + B K_{\nu}(pz)] e^{-ip\cdot x}
\eeqn
where 
\beq
    \nu =\bigl[ 4 + {m^2\over k^2} \bigr]^{1/2} \  .
\eeqn
For a spin-$\half$ field satisfying  the Dirac equation with mass $m$,
the solutions in Euclidean space have the form
\beqa
   \psi_L  &=& u_L(p)  z^{5/2}  [  A  I_{\nu_+} (pz) + B K_{\nu_+}(pz)] e^{-ip\cdot x} \CR
   \psi_R  &=& u_R(p)  z^{5/2}   [  A  I_{\nu_-} (pz) + B K_{\nu_-}(pz)] e^{-ip\cdot
     x}  \ , 
\eeqan
where
\beq
    \nu_\pm =  c \pm \half \ , \quad \mbox{with}  \quad c = {m\over
      k} . 
\eeqn
The parameter $c$ will play an important role in the physics discussed
in this paper.

For a spin-1 gauge field, using the background Feynman gauge of
Randall and Schwartz~\cite{RandallSchwartz}, the  solutions in Euclidean space have the form
\beqa
    A_m&=&   \eps_m(p) \   z^1 [ A  I_{1} (pz) + B K_{1}(pz)] e^{-ip\cdot x} \CR
     A_5 &=&      z^1 [ A  I_{0} (pz) + B K_{0}(pz)] e^{-ip\cdot x} \CR
    c &=&    z^1 [ A  I_{1} (pz) + B K_{1}(pz)] e^{-ip\cdot x}   \ ,
\eeqan
   where $c(x,z)$ is the ghost field.  The gauge boson system then
   mimics the system of a Dirac fermion with $c = \half$.   This
   correspondence allows us to compute the effects of gauge bosons
   by borrowing results from the fermionic case.   This fact and
   other relevant details of this construction are explained  in Appendix B.

By integrating out fields in the 5-dimensional bulk in the presence of
a fixed background gauge field, we generate an effective potential for
that gauge field, the Coleman-Weinberg potential.   The
Coleman-Weinberg potential is computed as an integral over Euclidean 4-momenta
\beq
     V =  \int { d^4p\over (2\pi)^4}  \left[
      -2  \log\det(\Psi) + {3\over 2} \log \det (A) \right],
\eeq{CWpot}
for the terms due to fermions and gauge fields. Precise expressions
for the operators labelled $\Psi$ and $A$ are given in Appendices A and B.  The
similarity of the solutions for these fields allows us to write a
general formula for the values of these determinants.   We emphasize
that, throughout this paper, we are not interested in the overall
constant in \leqn{CWpot} but only in the dependence on the Higgs
field, which appears here as a background $A_5^A$ field.

Consider, then, a field whose classical solutions take the form
\beq
     \Phi =   z^a [ A  I_\nu(pz) + B K_\nu(pz) ] e^{-ip\cdot x}
\eeqn
It is useful to define combinations of the Bessel functions with
definite boundary conditions at a point $z = z_2$,
\beq
    G_{\alpha\beta}(z_1,z_2) =  K_{\alpha}(p z_1)  I_{\beta} (pz_2)
   - (-1)^\delta I_\alpha(p z_1)  K_\beta (pz_2) \ , 
\eeq{bigGdef}
where $\alpha, \beta = \pm 1$, $(-1)^\delta = 1$ for $\alpha = \beta$
and $-1$ for $\alpha \neq \beta$, 
and the orders of the Bessel functions are 
\beq
  \mbox{for} \ \alpha,\beta =+1\  : \ \nu_+ =  c+\half \ ; \qquad 
 \mbox{for} \ \alpha,\beta =-1\  : \  \nu_- = c-\half
\eeqn
for an appropriate value of the parameter $c$.
Then $G_{++}(z,z_R)$, $G_{--}(z,z_R)$ will give solutions
 with Dirichlet boundary conditions
on the IR brane:  $\Phi(z,z_R) = 0$ at $z = z_R$. Due to the identities
\beq
         {d\over dz}  z^{\nu} I_\nu(z) = z^{\nu} I_{\nu-1}(z)  \quad  \mbox{and} \quad 
         {d\over dz}  z^{-\nu} I_\nu(z) = z^{-\nu} I_{\nu+1}(z), 
\eeqn
and similarly for other Bessel functions,
$G_{+-}(z,z_R)$, $G_{-+}(z,z_R)$ will give solutions with appropriate 
Neumann boundary 
conditions on the IR  brane.  The definition of Neumann boundary
conditions for gauge fields  and of both sets of boundary conditions 
for fermions requires some further explanation, which we give
in Appendices A and B.
   In this paper, we will refer to these Neumann and Dirichlet 
boundary conditions as $+$ and $-$ boundary conditions, respectively.

The $G$ functions obey the important identity
\beq
    G_{++}(z_1,z_2) G_{--}(z_1,z_2) -  G_{+-}(z_1,z_2) G_{-+}(z_1,z_2)
    =  -   {1\over p^2 z_1 z_2} \ ,
\eeq{Wronskian}
which follows from the Wronskian identity for Bessel functions.

We will use the $G$ functions to construct Green's functions for the
RS fields.  As an example, consider
\beq
     \eta_{mn} {\cal G}^{AB}(z,z',p)   =    \VEV{ A^A_m(z,p) A^B_n(z',-p)}   \ .
\eeq{Gfdef}
This object  is locally a solution of the classical
field
equations in $z$, satisfying three sets of boundary conditions.  These are:
(1) $+$ or $-$ boundary conditions on the IR brane at $z = z_R$,
(2) a discontinuity in the derivative of a fixed size at $z = z'$,
(3) $+$, $-$, or other appropriate boundary conditions at $z = z_0$.
For the field $A_m^A$, the  
solutions to the field equations will be a linear combination of 
\beq
        z^1    G_{++}(z,z_R) \quad \textrm{and} \quad     z^1  G_{+-}(z,z_R) \  ,
\eeqn
with $c=\half$. Take, for definiteness, Neumann boundary conditions at $z= z_R$. Then
the 
Greens function will have the form
\beqa
   {\cal G}^{AB}(z,z',p)   &=&    K  z z^{\prime} \CR
 & & \hskip -0.5in\cdot \cases{
         {\bf A}^{AB} G_{+-}(z,z_R) G_{+-}(z',z_R) -  \delta^{AB} G_{++}(z,z_R)
         G_{+-}(z',z_R)  &    $ z < z'$ \cr
     {\bf A}^{AB}  G_{+-}(z,z_R) G_{+-}(z',z_R) -  \delta^{AB} G_{+-}(z,z_R)
         G_{++}(z',z_R) &    $ z > z'$ \cr }  \ .
\eeqa{Gform}
In this formula, the second index $-$ insures the Neumann boundary
conditions at $z = z_R$. The constant $K$, which is independent of
species,
is determined by the discontinuity at $z = z'$. 
  The matrix ${\bf   A^{AB}}$, 
which depends on $p$ but is independent of $z$ and $z'$, is
  still undetermined at this stage.

To find ${\bf A}^{AB}$, we must fix the boundary condition at $z =
z_0$.  For example, if we directly apply Neumann boundary conditions at
$z = z_0$, we find the constraint
\beq
       \biggl[ {\bf A}^{AB} G_{--}(z_0,z_R)  - \delta^{AB} G_{-+}(z_0,z_R) \biggr] \
       G_{+-}(z',z_R) = 0 \ . 
\eeq{firstGbc}
The dependence on $z'$ factors away, as it must, and we find a simple
linear equation for ${\bf A}^{AB}$, leading to 
\beq
     {\bf A}^{AB} =  \delta^{AB} { G_{-+}(z_0,z_R) \over G_{--}
       (z_0,z_R) }  \ . 
\eeq{firstbcsol}
Imposition of the boundary condition at $z = z_0$ will always give us
a solution for ${\bf A}^{AB}$ in terms of functions \leqn{bigGdef}
evaluated at $(z_0, z_R)$, so, in the rest of this paper, we will
abbreviate
\beq
     G_{\alpha\beta} \equiv  G_{\alpha\beta}(z_0,z_R) \ .
\eeq{shortbigG}

In the examples of interest later in this paper, we will not apply
simple Dirichlet or Neumann boundary conditions directly to the
elementary fields.  Instead, we will apply these boundary conditions
only after the fields are mixed by a unitary transformation.  We will
explain in Section 3 how this allows us to encode the effect of the
Higgs field vacuum expectation value and other physical effects on the
UV brane.   In the presence of such a unitary transformation $U$, and
allowing more general boundary conditions, \leqn{firstGbc} is
generalized to the condition 
\beq
      U_{AC}  \biggl[ {\bf A}^{CB} G^{(C)}_{-A_0, -C_R}(z_0,z_R) 
 - \delta^{CB} G^{(C)}_{-A_0,+C_R}(z_0,z_R) \biggr] \
       G^{(B)}_{-B_0, -B_R}(z',z_R) = 0 \ . 
\eeq{secondGbc}
The notation of this equation is as follows:   $A_0$ represents the
boundary condition of the field $A$ at $z = z_0$.  That is,
$-A_0$ is $-$ if the field $A$ has $+$  (Neumann) UV boundary
conditions, and $+$ if the field has $-$ (Dirichlet) UV boundary
conditions.
The index $-C_R$  similarly reflects the IR boundary condition of the
field $C$.  In the gauge field case, the functions $G_{\alpha\beta}$
are fixed, but in the fermion case, these functions will depend on the
mass parameter $c$.  Since the twist $U$ is only on the UV brane, the
functions $G_{\alpha,\beta}$ in the bracket must be evaluated using the IR
identification of the field, that is, with the $c$ parameter of the
field $C$.  We denote this explicitly in \leqn{secondGbc}; the
superscript $(C)$ on  a $G$ function indicates that this function
should be evaluated with $c = c(C)$. 

The equation \leqn{secondGbc} is a linear equation for the matrix
${\bf A}^{AB}$. We   will now abbreviate this  equation as
\beq
         {\bf C}_{AC} \   {\bf A}^{CB} =    \mbox{(RHS)}  \ , 
\eeq{Cmatrixdef}
where 
\beq
   {\bf C}_{AC} =   U_{AC}  G^{(C)}_{-A_0, -C_R}  \ .
\eeq{Ceval}
The matrix $\CC$ depends on the 4-momentum $p$ through the $G$ functions
\leqn{bigGdef}, \leqn{shortbigG}.
Here we note explicitly that the indices $\nu$ of the Bessel functions in \leqn{bigGdef}
are to be evaluated using the IR field identification.   It will be
convenient to notate \leqn{Ceval} in a more abstract way as
\beq 
   {\bf C} =  \overrightarrow{B}_{UV}  U G
   \overleftarrow{B}_{IR} \ , 
\eeq{Cformula}
imagining that the operators $ \overrightarrow{B}_{UV}  $, $
\overleftarrow{B}_{IR} \ $  supply the appropriate UV and IR boundary conditions.

We are now ready to evaluate  the Coleman-Weinberg 
potential \leqn{CWpot}.   The determinants in this expression are
formally constructed as products over the KK mass spectrum 
\beq
         \det(A) =  \prod_i (p^2 + m_i^2)
\eeqn
The masses $m_i^2$ that appear in this formula can be identified as 
poles in the corresponding Green's functions.  So we must go back
through the solution for the Green's function given above and ask how
these poles could appear.   The Bessel functions in the
explicit factors of $G_{\alpha\beta}(z,z_R)$, $G_{\alpha\beta}(z',z_R)$
have no poles in
$p$, and the constant $K$ can be seen to be simply proportional to $p$.
Thus,
the poles must reside in ${\bf A}^{AB}(p)$, and must be generated when we invert the 
equation \leqn{Cmatrixdef}.  This observation implies 
{\it Falkowski's Theorem}~\cite{Falkowski} ,
\beq
    \det(A) =  \det  {\bf C}    \ , 
\eeq{FalkowskiThm}
where ${\bf C}(p)$ is the matrix in  \leqn{Cformula}, up to an overall
multiplicative constant. This constant  could in principle depend on
$U$, but it will be independent of $U$ if
$\det U = 1$.  With this identification, we reduce the
calculation of the functional determinant to the calculation of a
simple matrix determinant involving the functions $G_{\alpha\beta}$. 

In case this argument of   Falkowski for the identification
\leqn{FalkowskiThm} is not persuasive, we give a more constructive
argument for this result in Appendix C.

\section{Identification and influence of Higgs bosons}

Our next task is to define the UV boundary conditions on the fermion
and vector fields, and to review how these boundary conditions
incorporate
the effects of the Higgs boson vacuum expectation values.

In gauge-Higgs unification, the Higgs fields  arise as the 5th
components of gauge fields $A_M^A$.    These components transform as
scalars under
4-dimensional Lorentz transformations, so they can obtain a vacuum
expectation value.  Since it has nontrivial quantum numbers under the
gauge group, this expectation value can break down the gauge symmetry.

\begin{figure}
\centering
\includegraphics[width=4in]{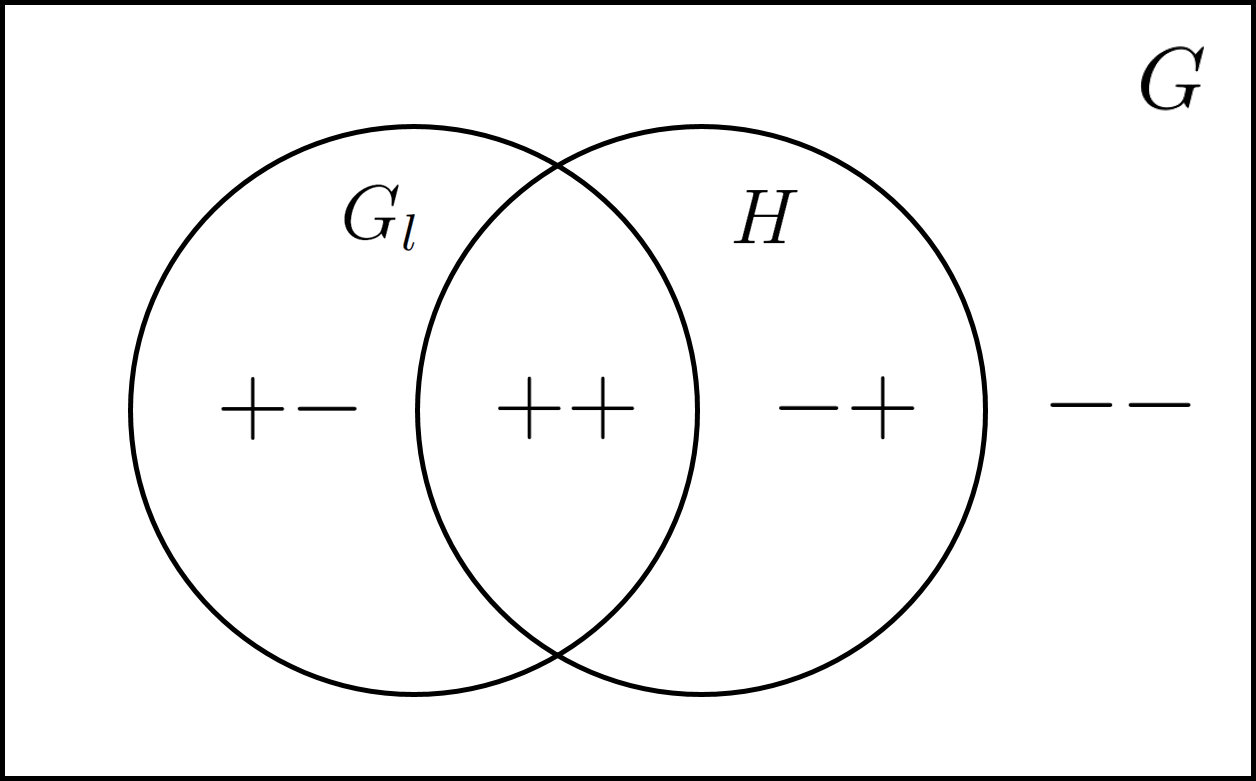}
\caption{The pattern of symmetry breaking. ($\pm \pm$) denotes boundary conditions of 5D gauge fields.}
\label{fig:Venn}
\end{figure}

We view the 5-dimensional theory as a dual description of a strongly
interacting 4-dimensional theory.  Our physical picture is that the strong
interaction theory has a global symmetry $G$ and a local 
gauge symmetry $G_\ell$ at the scale $1/z_0$.   The strongly
interacting theory spontaneously breaks the global symmetry $G$ to a
subgroup $H$ at the scale $1/z_R$.   This gives rise to the familiar
Venn
diagram shown in Fig.~\ref{fig:Venn}. 

 The gauge fields of the
5-dimensional
theory fit into this structure in different ways depending on their UV
and IR boundary conditions.   We will quote boundary conditions as $+$
or $-$ boundary conditions on $A^A_m$.  Because
boundary conditions are imposed on the gauge-covariant $F_{m5}^A$, 
\beq
              F_{m5}^A  =  \del_m A_5^A - \del_5 A_m^A + g f^{ABC}
              A_m^B A_5^C \ , 
\eeq{FofA}
a $+$ boundary condition for $A_m^A$ is only consistent with  a $-$
boundary condition for $A_5^A$ and vice versa.   In general, a field can have
{\it zero modes}, 
zero-energy solutions to the field equations, only with $++$ boundary
conditions.  Then $A_m^A$ will have zero modes for $++$ boundary
conditions and $A_5^A$ will have zero modes for $--$ boundary
conditions on $A_m^A$.   Zero modes in 5 dimensions are dual to
massless particles of the same 4-dimensional spin in 4 dimensions.

These considerations fit together into an  appealing picture.  Gauge
field components with $++$ boundary conditions give massless gauge
fields in 4 dimensons.   Gauge fields with $--$ boundary conditions give
massless scalars in 4 dimensions from $A_5^A$.   These will be
Goldstone bosons of the 4-dimensional theory.  Coming from the other
side of the duality, the underlying gauge symmetries of the 4-dimensional
theory can be identified with gauge  fields with $+$ UV boundary
 conditions, while ungauged generators of the global symmetry group
 are identified with gauge fields with $-$ UV boundary conditions.
Global symmetries that are broken at $1/z_R$ are identified with gauge
fields with $-$ IR boundary conditions, while unbroken global
symmetries are identified with gauge fields with $+$ IR
boundary
 conditions.   This correspondence, shown also in Fig.~\ref{fig:Venn},
 precisely identifies $++$ gauge fields with unbroken gauge symmetries
 and $--$ gauge fields with spontaneously broken global symmetries.

We now have a picture in which Higgs bosons appear as Goldstone bosons
of the symmetry breaking in the new strong interaction theory modelled by
the 5-dimensional RS fields.   This realizes  the idea of Higgs fields
as Goldstone bosons as proposed in \cite{GeorgiKaplan} and more
recently revived in the ``Little Higgs'' program~\cite{LHone,LHtwo}.
The little hierarchy is produced if the scale of the strong
interaction theory, associated with $1/z_R$, is much larger than the
Higgs field mass and vacuum expectation value.   To model this, we
take the RS setup as given and generate the Higgs potential from
radiative corrections to this picture, described quantitatively by the
Coleman-Weinberg potential.

We have presented a formalism for computing the Coleman-Weinberg
potential in the previous section.   How can a Higgs boson vacuum
expectation value be included? 

The Higgs bosons appear as  zero modes of fields $A_5^A$.
 A pure $A^A_5$ background field can always be 
removed locally by a gauge transformation.   However, in a
5-dimensional system with boundaries, the influence of $A^A_5(z,x)$
cannot be gauged away completely.   There is gauge-invariant
information
parametrized by the Wilson line
\beq
     W[A] =   P \biggl\{ \exp\bigl[  i g \int_{z_0}^{z_R} dz \, A_5^A
    T^A\bigr]\biggr\} \  .
\eeq{Wilson}
The Coleman-Weinberg potential can depend on the Wilson line and, in
this way, on the expectation value of $A^A_5$. 

In the formalism of the previous section, the Wilson line appears in
the following way:  The equations in the previous section apply to
free fermion and gauge fields with zero background Higgs fields.
We can apply these same formulae to a system with a background $A_5^A$
field if we gauge away $A_5^A$ in the central region of $z$,
leaving a singular field near $z_0$ or $z_R$.   The effect of a
nonzero $A_5^A$ field is
implemented by applying the Wilson line as a matrix to the various fields in the
problem,  setting  $T^A = t^A$, the representation matrix 
in the appropriate representation of
the gauge group $G$.  In this paper, we will generally consider the $A_5^A$
field as gauged away to the UV boundary.  (It is a check on our
formalism that the same results can be obtained by gauging away
$A_5^A$ to the IR boundary.)  

The zero mode of $A_5^A$, present when the $A_m^A$ field has boundary
conditions $--$, has the form 
\beq
                   A_5^A(z,x^m) =  N_h \ z \ h^A(x^m)    \ , 
\eeq{bkgdfield}
where the $z$ dependence is that of the $A_5^A$ zero mode 
and $N_h$ is a normalization constant.   Then let
\beq
      U_W =   \exp\bigl[ - i g \int_{z_0}^{z_R} dz \,  N_h z \VEV{h^A}t^A\bigr]
\eeq{UWdef}
The matrix $U_W$ should be applied to each field before imposing the
boundary condition at $z = z_0$.    In this context, the matrix $U_W$
plays the role of the matrix $U$ in \leqn{Cformula}.

There may be additional complications that influence the UV boundary
conditions.  For example, it is allowed to introduce a fermion mass
term on the boundary,
\beq
   \delta L =  \sum_{ij}  M_{ij} \bar\Psi_i \Psi_j \  
 \delta(z - z_0) \   ,
\eeq{massonboundary}
as long as the mass matrix $M_{ij}$ preserves the assumed local  gauge
symmetry by mixing only fermion fields with the
same $G_\ell$ quantum numbers. In this paper, we will include such a
mass mixing only on the UV boundary.   The effect of this term in models will be
to mix fermions actively participating in electroweak symmetry
breaking with  the light quarks and leptons, similarly to the Extended
Technicolor interaction.  The influence of
\leqn{massonboundary} is to mix the fermion fields by a unitary
transformation.   We will implement this directly by including a
unitary matrix $U_M$ before applying the UV boundary condition.

Our final expression for the matrix ${\bf C}$ is 
\beq 
   {\bf C}=  \overrightarrow{B}_{UV} U_M U_W G
   \overleftarrow{B}_{IR} \ . 
\eeq{UUCformula}

This formula has an important property that we will use often in our
discussion.  If fermions mixed by $U_M$ have the same boundary
condition in the UV, then    $[U_M, \overrightarrow{B}_{UV}] = 0$.   Then
we can move $U_M$ to the left and find
\beq
    \det {\bf C} =     \det U_M \cdot  \det \biggl[  
\overrightarrow{B}_{UV} U_W G
   \overleftarrow{B}_{IR} \biggr] \ . 
\eeq{moveUleft}
Since $U_M$ is unitary, $\det U_M  = 1$, and the mixing  angles in 
 $U_M$ disappear from the expression for the Coleman-Weinberg
 potential.
Similarly, if $U_W$ mixes only fields with the same IR boundary
conditions, we can move $U_W$ to the right of  
$\overleftarrow{B}_{IR} $ and factor it out of determinant
calculation.  Since $\det U_W = 1$, the mixing angles in $U_W$ do not
contribute to the Coleman-Weinberg potential.  The Higgs field appears
as mixing angles in $U_W$ and so, in this latter case, the
Coleman-Weinberg potential is flat in $\VEV{h}$. 

This argument extends to decompositions of $U_W$ and $U_M$: if 
 \beq
    U_MU_W =  U_1 U \ \mbox{and} \   [U_1,
    \overrightarrow{B}_{UV}] = 0 
\eeq{dropright}
then the Coleman-Weinberg potential does not depend on the angles in $U_1$.
In moving pieces of the unitary matrix to the right, we must be more
careful. The matrix $U_W$ mixes fermions within a gauge multiplet, and
these must have the same values of $c$, but $U_M$ generally mixes fermions
in different multiplets with different values of $c$.  
The fermion Green's function $G(z_0,z_R)$ depends on $c$.   So, if 
\beq
   U_W =  U U_2  \ \mbox{and} \   [U_2,
    \overleftarrow{B}_{IR}] = 0 
\eeq{dropleft}
then  $U_2$ does not contribute to the Coleman-Weinberg potential.
More generally, pieces of $U_M$ may be moved to the right and
eliminated if they mix fermion fields with the same value of $c$.

\section{Fermion zero modes}

Just as the boundary conditions on gauge fields have physical
significance, the boundary conditions on fermion fields have a
significance for model-building.   Five-dimensional fermions are 
4-component Dirac fermions, but, with appropriate boundary conditions,
they can have zero modes that can be interpreted as chiral quarks and
leptons~\cite{Grossman,GP}.

The zero-mode solutions of the Dirac equation are present for any
nonzero value of the 5-dimensional fermion mass.  With, again,
\beq
           c = m/k \ , 
\eeq{cdefin}
a zero mode corresponding to a left-handed 4-dimensional fermion has
the form 
\beq
    \psi_L = f_-  u_L(p) z^{2-c} e^{-ip\cdot x}   \qquad   \psi_R  =
    0   \  ,
\eeq{leftzero}
where $u_L(p)$ is the usual 2-component massless spinor of a
left-handed fermion and $f_-$ is a normalization constant.   Similarly, 
a zero mode corresponding to a right-handed 4-dimensional fermion has
the form 
\beq
    \psi_R=  f_+ u_R(p) z^{2+c} e^{-ip\cdot x}   \qquad   \psi_L  =
    0 \   .
\eeq{rightzero}
We will refer to these as $L$ and $R$ zero modes, respectively.
These zero modes are nonzero at the boundary, and so they require
appropriate fermion boundary conditions,  $++$ for the L zero mode and
$--$ for the R zero mode.  

An important feature of the zero modes is their structure in the 5th
dimension.
The probability distribution of the position in the 5th dimension is
given, for the L zero mode, by 
\beq
   \int dz \sqrt{g} \bar \Psi (kz \gamma^0) \Psi   =  \int dz { kz \over
     (kz)^5} \bigl| f_-\bigr|^2   z^{4-2c} \sim  \int {dz \over z}
   z^{1-2c}  \ .
\eeq{zprobdist}
For $c > \half$, the zero mode is localized near the UV brane; for $c
< \half$, the zero mode is localized near the IR brane.   For the $R$
zero mode, the same calculation gives the boundary at $c = - \half$.
Again,
\beq
  \begin{tabular}{lccc}
   &    $ c < -\half$   \quad  &  \quad $  -\half < c < \half $  \quad  &  \quad $ \half < c $  \\
    \\
L \quad  &      IR     \quad       &  \quad  IR      \quad     & \quad UV \\
R  \quad &      UV       \quad    &  \quad  IR      \quad    &   \quad IR  \\
\end{tabular} \ .
\eeq{UVIRtable}
       
In a realistic model, the light quarks and leptons would be described
by UV zero modes.   We will see in a moment that the formation of a 
symmetry-breaking potential probably requires a pair of IR zero modes
which are mixed by a symmetry-breaking Higgs expectation value.   The
right-handed top quark can potentially be assigned to an IR zero mode.
The assignment of the left-handed $(t,b)_L$ doublet to IR zero modes
is 
potentially in tension with precision electroweak
 constraints on the $b_L$.  This is
an important issue for model-building~\cite{ADMSundrum}.

\section{Simplest examples of symmetry breaking}

As a first application of this formalism, we review the calculation of
the Higgs potential from one fermion multiplet by Contino, Nomura, and
Pomarol~\cite{CNP}.   We will do this in the simplest context of an
$SU(2)$ gauge field acting on two-fermion multiplets.   We assign 
the boundary conditions for the two fermion fields as
\beq
   \Psi \sim    \pmatrix{ + & + \cr - & - \cr} \ .
\eeq{firstbc}
The notation here is to write the UV and IR  boundary conditions,
respectively, for each fermion component on a horizontal line.  The
matrix represents a single $G$ representation.
Fermions in the same $G$ representation  must have the same value of
$c = m/k$, the parameter that determines the localization of the zero
modes.    Consistently with these assignments, the gauge fields must
be assigned boundary conditions that break $SU(2)$ down to its $U(1)$
subgroup,
\beq 
      A^A_m  \sim   \pmatrix {   - & - \cr   - & - \cr  + & + \cr} \,
\eeqn
for $A^1_m$, $A^2_m$, $A^3_m$, respectively.   Note that, with these
assignments, $A_5^1$ and $A_5^2$ are Goldstone bosons.

Now turn on $\VEV{A_5^2 }\neq 0$.  This is a direction that breaks the
$U(1)$ gauge symmetry and mixes the two fermion
components.   The
corresponding $U_W$ is 
\beq
      U_W =   \pmatrix{  c_W  &    -s_W\cr  s_W
        & c_W \cr} \ , 
\eeq{firstUW}
where $c_W = \cos\theta$, $s_W = \sin\theta$, with 
\beq 
       \theta =  {  g\over 2}  \int_{z_0}^{z_R}  dz \   A_5^2(z)  \ .
\eeq{firsttheta}

Combining \leqn{firstbc} and \leqn{firstUW}, we find the $\CC(p)$
matrix from \leqn{Ceval} or \leqn{Cformula} as 
\beq
       \CC =    \pmatrix{ c_W G_{--}&  -s_W
           G_{-+}\cr s_W G_{+-}  &  c_W
           G_{++} \cr } \ .
\eeq{firstCmatrix}
We find immediately
\beqa
  \det \CC &=&   c_W^2   G_{--} G_{++} + s_W^2 G_{-+} G_{+-} \CR
   &=&   (G_{--} G_{++}) \biggl(   1  -  s_W^2   ( G_{--} G_{++} - G_{-+} G_{+-})/
 G_{--} G_{++} \biggr)\ .
\eeqa{firstdet}
The first factor is independent of $\theta_W$, so we can ignore it.
The second factor simplifies with the use of the identity
\leqn{Wronskian}, which can be abbreviated here as 
\beq
  G_{--} G_{++} - G_{-+} G_{+-}   = -  { 1\over  p^2  z_0 z_R}\ .
\eeq{Wronskiantwo}
We then find 
\beq
  \log   \det \CC =  \log \biggl[  1  + { s_W^2  \over  p^2 z_0 z_R
       G_{--} \  G_{++}} \biggr] \ .
\eeq{firstlog}

In the Euclidean region, for $z_0 \ll z_R$, all four Green's functions 
$ G_{--},  G_{-+}, 
     G_{++},   G_{+-}$ are positive definite functions of $p$.
All four functions increase exponentially for large $p$, as 
\beq
      G_{ab}(z_0,z_R) \sim  e^{ p(z_R - z_0) } \ .
\eeq{Gasympt}

\begin{figure}
\centering
\includegraphics[width=4in]{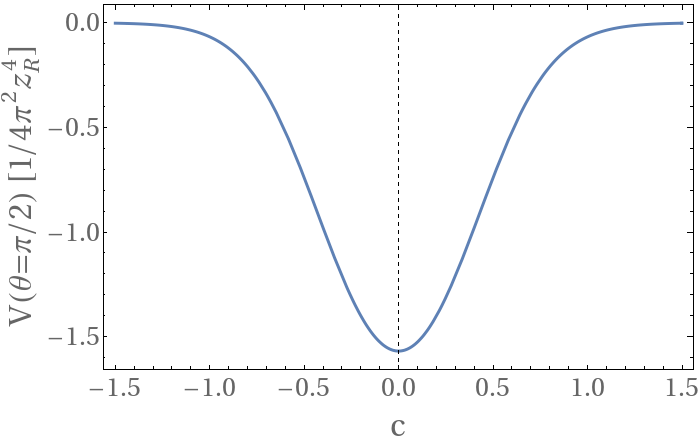}
\caption{Dependence of the  depth of the minimum of the
  Coleman-Weinberg potential
\leqn{firstfinal} on the parameter $c$.}
\label{fig:cpotential}
\end{figure}

The fermionic contribution to the Coleman-Weinberg potential for this
model is
then~\cite{CNP}
\beq
    V(h) =  - 2 \int{d^4 p\over (2\pi)^4}  \log \biggl[ 1 + { s_W^2
    \over   p^2  z_0 z_R    G_{--}  G_{++} }
  \biggr] \ . 
\eeq{firstfinal}
This result is well-defined and UV convergent and is negative
definite.   It is minimized at $\theta = \pi/2$.    The depth of the
potential  depends strongly on the parameter $c$, as shown in 
Fig.~\ref{fig:cpotential}.
The finiteness of the Coleman-Weinberg potential is an important 
 general feature
of  gauge-Higgs unification models.   It follows from the fact that
the Wilson line order parameter of the symmetry breaking is a nonlocal
quantity.
Since $\VEV{A_5^A}$ can be gauged away locally, the potential does
not get contributions from the deep ultraviolet.  However, the energy
scale of the potential is set by $z_R$, and so there is still
a little hierarchy if $1/z_R \gg 100$~GeV.

We must add to the fermion contribution the result for the
Coleman-Weinberg potential of the vector bosons.   The same formalism
applies.
The $SU(2)$ gauge group acts on the three gauge boson fields $A^1_M,
A^2_M, A^3_M$ according to 
\beq
     t^a_{bc}  = i \eps^{bac}   \ .
\eeqn
Then the  $U_W$ matrix for the three gauge boson states is
\beq
       U_W = \pmatrix{ c_{2W} & 0 &  s_{2W}\cr  0 & 1 & 0 \cr
                 -  s_{2W} & 0 &  c_{2W}\cr }
\eeq{UWforgauge}
where  $c_{2W} = \cos 2\theta$, $s_{2W} = \sin 2 \theta$, where $\theta$ is
as in \leqn{firsttheta}.    The boundary conditions on the fields
$A_M^1$ and $A_M^3$ are the same as those on the two fermion fields in
this example.  Then we find that the Coleman-Weinberg potential
generated by the gauge fields is
\beq
    V(h) =  +{3\over 2} \int{d^4 p\over (2\pi)^4}  \log \biggl[ 1 + { s_{2W}^2
    \over   p^2 z_0 z_R    G_{--}  G_{++} }
  \biggr] \ . 
\eeq{firstfinalV}
where, in this expression, $G_{--}$ and $G_{++}$ are evaulated at  $c = \half$.

\begin{figure}
\centering
\includegraphics[width=4in]{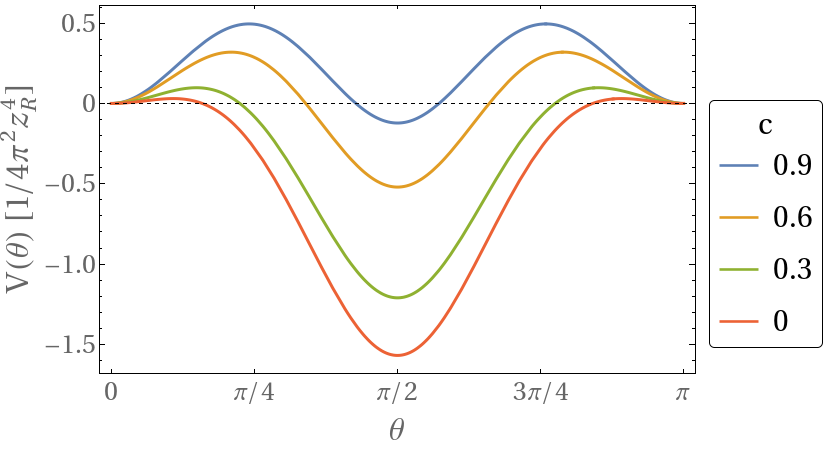}
\caption{The complete Coleman-Weinberg potential for the model of
  Section~5,  including both fermion and gauge boson contribution.}
\label{fig:firstcomplete}
\end{figure}

Some  graphs of the complete potential for this model, with $c =
\half$ for the gauge fields and different values of $c$ for the
fermions, are shown in Fig~\ref{fig:firstcomplete}.  In the typical
situation, there is a potential barrier between the symmetric point at
$h = 0$ and the symmetry-breaking minimum; that is, the phase
transition is first-order and it is not possible to tune the value of
$v/f$ to be small.   This is still true if the number of fermion
flavors
is taken to be a variable $n_f$ and varied continuously.   The minimum
of the potential is always either at $\theta = 0$ or $\theta = \pi$. 

\section{Competing forces with 2 fermion multiplets}

To incorporate a  little hierarchy with $v/f \ll 1$, a model must be
in the vicinity of a second-order phase transition in the space of
parameters of the Coleman-Weinberg potential.  We would like to
understand systematically how to achieve this in models with multiple
fermion and gauge fields.    In this
section, we take a first step into this program by working out the
possible phase diagrams of systems of two fermion multiplets.   For simplicity,
we will restrict ourselves to $SU(2)$ in this section, and we will
ignore the 
gauge field contributions to the potential.    We call the two fermion
multiplets
$\psi_1$ and $\psi_2$ and assign them mass parameters $c_1$ and $c_2$.
We will call the  Green's functions associated with these multiplets
$G^1_{ab}$ and $G^2_{ab}$, respectively.  An example of the full
expansion of this  notation is
\beq 
    G^1_{+-} =  G^{(\psi_1)}_{+-}(z_0,z_R) \ . 
\eeq{Goneexample}

\subsection{No UV mixing}

Consider first the simplest case in which there is no UV mass mixing
($U_M = {\bf 1}$).    In this case, we have a pair of fermion
representations in the {\bf 2} of $SU(2)$, with boundary conditions
such as
\beq
            \pmatrix{ + & + \cr - & - \cr} \qquad  \pmatrix{ + & - \cr - & - \cr} 
\eeqn
In Appendix A, we show that the reversal of the $c$ parameter and the
UV and IR boundary conditions
\beq
           c\to -c \ ,  \qquad      + \leftrightarrow  - 
\eeq{reversal}
is a symmetry of a free fermion in RS.  According to the argument
given below \leqn{moveUleft}, a fermion multiplet gives zero
contribution to the Coleman-Weinberg potential if either its two UV
boundary
conditions or its two IR boundary conditions are identical.  Then, for
one fermion multiplet, 
there are only two possible situation in which we obtain a nonzero
Coleman-Weinberg potential
 \beq
        \psi_A \sim    \pmatrix{ + & + \cr - & - \cr} \qquad
        \mbox{and} \qquad \psi_R \sim  \pmatrix{ + & - \cr - & + \cr}
        \ . 
\eeq{twocases}
We call these the A (``attractive'') and R (``repulsive'') cases,
respectively.  

The potential in the attractive case was worked out in
\leqn{firstfinal} above.  For future reference, we notate this
potential as a function of  $s_W = \sin\theta$ and the fermion mass
parameter $c$, 
\beq
    V_A(s_W,c ) =  - 2 \int{d^4 p\over (2\pi)^4}  \log \biggl[ 1 + { s_W^2
    \over  p^2  z_0 z_R    G_{--} G_{++} }
  \biggr] \ .
\eeq{VAdef}
The potential is negative definite, and its minimum is always at $s_W
= 1$.

The potential in the repulsive case can be worked out in the same
way.  From 
 \beq
         \CC  =    \pmatrix{ c_W G_{-+}   &  -s_W
           G_{--}\cr s_W G_{++}   &  c_W
           G_{+-} \cr } \ ,
\eeq{CRmatrix}
  we find using the same method
\beq
    V_R(s_W,c ) =  - 2 \int{d^4 p\over (2\pi)^4}  \log \biggl[ 1 - { s_W^2
    \over   p^2  z_0 z_R    G_{-+} G_{+-} }
  \biggr] \ .
\eeq{VRdef}
This potential is positive definite, and its minimum is always at $s_W
= 0$. One way to understand its repulsive nature is that here the Higgs expectation value mixes two massive states and therefore lowers the mass of the lightest state. This is energetically less favored, so the potential resists forming a condensate.

\begin{figure}
\centering
\includegraphics[width=4in]{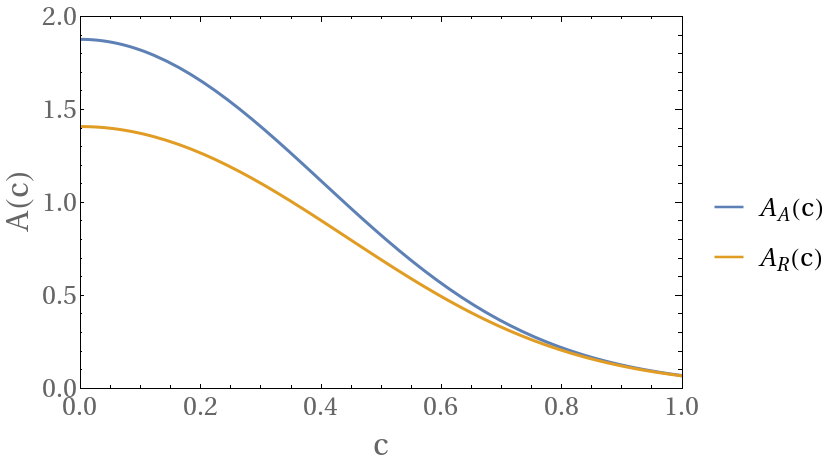}
\caption{The $c$-dependence of $A_A(c)$ and $A_R(c)$.}
\label{fig:Aline1}
\end{figure}

\begin{figure}
\centering
\includegraphics[width=4in]{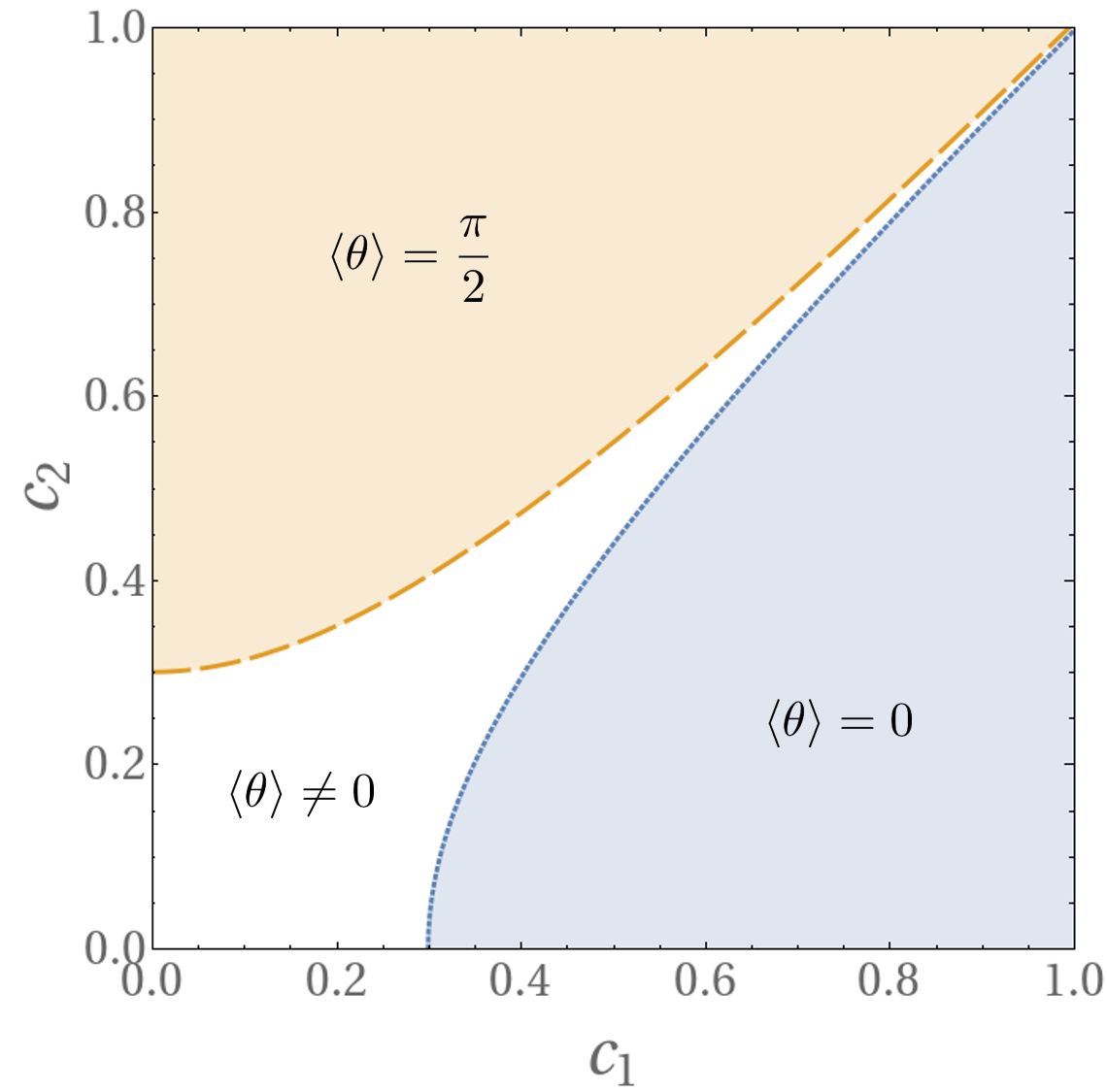}
\caption{Phase diagram of the model of Section 6.1 in the $c_1-c_2$ plane. In most values of $c_1$ and $c_2$, the minimum is at either $\VEV{\theta}=0$ or $\VEV{\theta}=\pi /2$. However, a non-trivial minimum is realized in the middle white area. Note that along the line $c_1=c_2$, $\VEV{\theta} = \pi /4$.}
\label{fig:Aline}
\end{figure}

If we have two fermion multiplets, one with the A type and one with
the R type boundary conditions, these two multiplets will compete.  To
understand the competition, we need to work out the expansions of
$V_A$ and $V_R$ about $s_W = 0$.  This is done in Appendix D.  
These expressions have expansions
in $\sin\theta$ with the forms
\beqa
      V_A(s_W,c) &= &  {1 \over 4 \pi^2 z_R^4 } \left[ -   A_A(c) s_W^2 +  {1 \over 2} B_A(c) s_W^4 +
     {1 \over 2} C_A(c) s_W^4\log (1/s_W^2)  + {\cal
        O}(s_W^6) \right] \CR
      V_R(s_W,c) &=& {1 \over 4 \pi^2 z_R^4 } \left[ +  A_R(c) s_W^2 + {1 \over 2} B_R(c) s_W^4 + {\cal
        O}(s_W^6) \right] \ ,
\eeqa{Vexpand}
where we have chosen the signs so that all of the coefficients are
positive functions of $c$.   Fig.~\ref{fig:Aline1} shows that $A_A(c) > A_R(c)$, but
both functions are rapidly decreasing functions of $c$.   Then there
is a line in the $(c_1,c_2)$ plane, shown as a dotted line in Fig.~\ref{fig:Aline}, where 
\beq
         A_A(c_1) = A_R(c_2) \  .
\eeq{Aline}
For $c_1$ slightly outside this boundary, the potential $V_A(c_1) +
V_R(c_2)$ has a negative quadratic term in $s_W$ that goes to zero on
the curve \leqn{Aline},  and a positive
quartic term.  Then the curve \leqn{Aline} is a line of second-order
transitions.  Near this line, the minimum $v$ of the potential can be made as
small as we like.  For the representative case $z_0/z_R = 0.01$, the tip of the curve occurs with  $c_2 = 0$ at
$c_1 =  0.2997$.

Actually, there are two minima, at $v$ and $-v$.
These minima merge to a single minimum at $\VEV{\theta} = \pi/2$ along the
line of bifurcations indicated by the dashed line in
Fig~\ref{fig:Aline}. 

\subsection{Cases with UV mixing}

In the remainder of this section, we will extend this analysis to the
more general case of two $SU(2)$  fermion multiplets with mass mixing
on the UV brane.   We will analyze the cases systematically for all
possible choices of fermion boundary conditions.  It is interesting,
at least to us, that all of the cases that we will encounter can be
understood from the competition  between attractive and repulsive
boundary conditions that we have seen already in Section 6.1.   That
is, this concept is  robust with respect to turning on fermion mixing
on the UV boundary.  In most cases, the generalization is relatively
straightforward, although the last case considered in Section 6.6 has some
nontrivial features.

A given fermion multiplet has $2^4$ possible boundary conditions, so a
pair of multiplets has $256$ different boundary conditions to
analyze.  However, many of these are related by the symmetry $+
\leftrightarrow -$, $c \leftrightarrow -c$ or by interchange of the
top and bottom components of the fermion multiplets.   Also, as we
showed at the end of Section 3,  mass
mixing on the UV brane has physical effect only if the fermions mixed
by the mass term have different UV boundary conditions.   The strategy
of our analysis will be to enumerate the different  possible IR boundary
conditions  and then to go through the 16 cases of UV boundary
conditions from the simplest to the most difficult, using symmetries
whereever possible to reduce cases to equivalent ones.

The 16 possible IR boundary conditions can be reduced to four cases.
Case I includes the two cases
\beqa
    \pmatrix{ \phantom{+} & + \cr
                              \phantom{+} & + \cr} & \qquad& 
\pmatrix{ \phantom{+} & + \cr
                              \phantom{+} & + \cr} \CR
   \pmatrix{ \phantom{+} & + \cr
                              \phantom{+} & + \cr} & \qquad& 
\pmatrix{ \phantom{+} & - \cr
                              \phantom{+} &- \cr} 
\eeqa{caseone}
and two more cases with $+
\leftrightarrow -$.   Case II includes the eight cases equivalent to 
\beq
    \pmatrix{ \phantom{+} & + \cr
                              \phantom{+} & - \cr}  \qquad
\pmatrix{ \phantom{+} & + \cr
                              \phantom{+} & + \cr} 
\eeq{casetwo}
Case III includes 
\beq
    \pmatrix{ \phantom{+} & + \cr
                              \phantom{+} & - \cr} \qquad
\pmatrix{ \phantom{+} & - \cr
                              \phantom{+} & + \cr} 
\eeq{casethree}
and the equivalent case with $+
\leftrightarrow -$.    Case IV includes 
\beq
    \pmatrix{ \phantom{+} & + \cr
                              \phantom{+} & - \cr} \qquad
\pmatrix{ \phantom{+} & + \cr
                              \phantom{+} & - \cr} 
\eeq{casefour}
and the equivalent case with $+
\leftrightarrow -$. 

For each case, we have 16 choices of UV boundary conditions. We will
also introduce mixing by angles  $\alpha$ between the two fermions in
the top row and $\beta$ between the two fermions in the bottom row.
In cases
in which the two multiplets have the same boundary conditions, such as 
\beq
    \pmatrix{+ &  \phantom{+} \cr
                          - &    \phantom{+} \cr} \qquad
\pmatrix{ + & \phantom{+} \cr   - &     \phantom{+} \cr} 
\eeq{leftbc}
   the mixing can be removed.   In cases such as 
   \beq
    \pmatrix{+ &  \phantom{+} \cr
                          - &    \phantom{+} \cr} \qquad
\pmatrix{ - & \phantom{+} \cr   - &     \phantom{+} \cr} \qquad
\mbox{and} \qquad
    \pmatrix{- &  \phantom{+} \cr
                          - &    \phantom{+} \cr} \qquad
\pmatrix{ + & \phantom{+} \cr - &     \phantom{+} \cr} 
\eeq{secondleftbc}
the mixing by $\beta$ has no effect but the potential depends on the
mixing angle $\alpha$. Actually, the two cases shown in
\leqn{secondleftbc} are equivalent, since increasing $\alpha $ by
$\pi/2$ interchanges the two boundary conditions in the top line.  So,
for each case listed above, we have a trivial situation in which the
potential is independent of $\alpha$ and $\beta$, situations in which
the potential only depends on one angle, and one case of the greatest
complexity in which the potential depends on both angles.
                  
\subsection{Case I}

In case I, the IR boundary conditions are the same for both fermions
in each multiplet.  Then, by the argument at the end of Section 3, the
Coleman-Weinberg potential is independent of $s_W$.  For all of these
cases, 
\beq
         V(s_W, c_1, c_2) = 0 \  .
\eeq{caseoneV}

\subsection{Case II}

For case II, we begin from the IR boundary conditions in \leqn{casetwo} and add 
UV boundary conditions, for which there are 16 possibilities.  These
can be grouped into three sets. 

  In the first set (4 cases), the UV boundary
conditions are the same between the two multiplets, and the
calculation of the Coleman-Weinberg potential reduces to that of two
separate multiplets.  For example, in 
\beq
    \pmatrix{+ & +\cr
                          - &   - \cr} \qquad
\pmatrix{ + & +\cr   - &    + \cr} 
\eeq{firstintwo}
the second multiplet gives zero and the combined potential is
obviously
\beq
    V(s_W,c_1,c_1) =  V_A(s_W,c_1) \ . 
\eeq{Vforfirstintwo}
All of these cases give a potential equal to either $V_A(s_W,c)$ or $V_R(s_W,c)$.

In the second set, with one pair of UV boundary conditions identical,
there are 8 cases, connected in pairs by $\alpha\to \alpha+\pi/2$ or 
$\beta \to\beta+\pi/2$.  An example is 
\beq
    \pmatrix{+ & +\cr
                         + &   - \cr} \qquad
\pmatrix{ -& +\cr   + &    + \cr} \ , 
\eeq{secondintwo}
for which the potential depends on $\alpha$ but not on $\beta$.   In
addition, the contribution to the potential from $\psi_2$ has no
dependence on $s_W$.  We can thus reduce the unitary transformation at
the UV brane to 
\beq 
    U   = \pmatrix{ c_\alpha & 0 & -s_\alpha & 0 \cr
                            0 & 1 & 0 & 0 \cr   s_\alpha & 0 & c_\alpha & 0 \cr
                            0 & 0 & 0 & 1 \cr } \pmatrix{  c_W & -s_W
                            & 0 & 0 \cr s_W & c_W & 0 & 0 \cr 0 & 0 &
                            1 & 0 \cr 0 & 0 & 0 & 1 \cr} = 
 \pmatrix{  c_\alpha c_W & - c_\alpha s_W & - s_\alpha & 0    \cr
  s_W & c_W & 0 & 0    \cr  s_\alpha c_W & - s_\alpha s_W & c_\alpha &
  0    \cr
  0 & 0 & 0 & 1 \cr}\ ,
\eeq{firstreduced}
where $c_\alpha = \cos\alpha$, $s_\alpha = \sin\alpha$. 
The matrix $\CC$ then has the form
\beq
 \CC = \pmatrix{  c_\alpha c_W G^1_{--} & - c_\alpha s_W G^1_{-+}& - s_\alpha   G^2_{--}  & 0\cr
  s_W G^1_{--}& c_W  G^1_{-+} & 0 & 0    \cr  s_\alpha c_W G^1_{+-}  &
  - s_\alpha s_W G^1_{++} & c_\alpha G^2_{+-} &
  0    \cr
  0 & 0 & 0 & G^2_{--}\cr}\   .
\eeq{Cforcasetwoone}
It is easy to check that $\det U = 1$.  The cancellations in
that calculation guide the simplification of $\det \CC$.  
We find
\beqa
\det \CC &= &(c_\alpha^2 G^1_{--}G^2_{+-} + s_\alpha^2 G^1_{+-} G^2_{--})
G^1_{-+} G^2_{--}  \CR & & \hskip 0.2in + s_\alpha^2 s_W^2 (G^2_{--})^2  (G^1_{--} G^1_{++}
- G^1_{-+}G^1_{+-}) \ .
\eeqa{detCforcasetwoone}
We can reduce the last term  using the Wronskian identity
\leqn{Wronskian} and extract a factor independent of $s_W^2$.    Then
the Coleman-Weinberg potential becomes
\beq
  V(s_W, c_1,c_2) = - 2 \int{d^4p\over (2\pi)^4} \log \biggl[ 1 -
    {s_\alpha^2 s_W^2 \over p^2 z_0 z_R G^1_{-+}G^1_{+-}} { 1\over
     ( s_\alpha^2 + c_\alpha^2 G^1_{--} G^2_{+-}/G^2_{--} G^1_{+-})}\biggr]
   \ . 
\eeqn
This  is a purely repulsive potential with strength diminished by
$s_\alpha^2$.  In fact, for $c_1 = c_2$, 
\beq
    V(s_W, c_1,c_2)  =   V_R(s_\alpha s_W, c_1) \ . 
\eeq{Vfortwoequal}

A way to guess the answer \leqn{Vfortwoequal} is to note that, for
$s_\alpha =0$, there is no mixing and the potential can be seen by
inspection to be zero, while for $s_\alpha = 1$, the UV boundary
conditions $+$ and $-$ in the top lines of \leqn{secondintwo} are
reversed and the potential is exactly that of the repulsive case in
Section 6.1.

The other three similar cases can be analyzed in the same way.  They
are either purely attractive or purely repulsive.   We quote the
results for the potential in the case $c_1 = c_2$:
\beqa
     \pmatrix{+ & +\cr
                         - &   - \cr} \quad
\pmatrix{ -& +\cr   - &    + \cr}  \quad  \to    \quad  V &=&  V_A( c_\alpha s_W, c_1) \CR
     \pmatrix{+ & +\cr
                         + &   - \cr} \quad
\pmatrix{ +& +\cr   - &    + \cr}   \quad   \to   \quad   V &=&  V_A( s_\beta s_W, c_1) \CR
   \pmatrix{- & +\cr
                         + &   - \cr} \quad
\pmatrix{ -& +\cr   - &    + \cr}   \quad   \to   \quad   V &=&  V_R( c_\beta
s_W,c_1) \ .
\eeqan

Finally, we come to the case in which the potential depends on both
mixing angles
\beq
     \pmatrix{+ & +\cr
                         + &   - \cr} \quad
\pmatrix{ -& +\cr   - &    + \cr}   \ . 
\eeq{casetwolast}
For this case, $U_M$ depends on both $\alpha$ and $\beta$, but 
we can still simplify $U_W$ as in \leqn{firstreduced},
so that 
\beq 
    U   = 
 \pmatrix{  c_\alpha c_W & - c_\alpha s_W & - s_\alpha & 0    \cr
 c_\beta s_W & c_\beta c_W & 0 & -s_\beta    \cr  s_\alpha c_W & - s_\alpha s_W & c_\alpha &
  0    \cr
 s_\beta s_W & s_\beta c_W & 0 & c_\beta \cr}\ .
\eeq{tworeduced}
The corresponding $\CC$ matrix is
\beq
\CC =  \pmatrix{  c_\alpha c_W G^1_{--}& - c_\alpha s_W G^1_{-+} & - s_\alpha G^2_{--}& 0    \cr
 c_\beta s_W G^1_{--} & c_\beta c_W G^1_{-+} & 0 & -s_\beta  G^2_{--}
 \cr  s_\alpha c_W G^1_{+-} & - s_\alpha s_W G^1_{++} & c_\alpha G^2_{+-} &
  0    \cr
 s_\beta s_W G^1_{+-} & s_\beta c_W G^1_{++} & 0 & c_\beta G^2_{+-} \cr}\ .
\eeqn
Then 
\beqa
\det \CC &=&  c_\alpha^2  c_\beta^2   G^1_{--} G^1_{-+} (G^2_{+-})^2 +
c_\alpha^2 s_\beta^2 G^1_{--} G^1_{++} G^2_{--} G^2_{+-} \CR
 & & + s_\alpha^2 c_\beta^2 G^1_{+-} G^1_{-+} G^2_{--} G^2_{+-} +
s_\alpha^2  s_\beta^2  G^1_{+-} G^1_{++} (G^2_{--})^2   \CR
& &  - (s_\alpha^2 c_\beta^2 - s_\beta^2 c_\alpha^2) s_W^2 G^2_{--}
G^2_{+-} / p^2 z_0 z_R  \ . 
\eeqan
We then find
\beq
 V = - 2 \int {d^4p\over (2\pi)^4}  \log \biggl[ 1 -   {s_-s_+
   s_W^2\over p^2 z_0 z_R  {\bf D} } \biggr]\ , 
\eeqn
where 
\beq
     s_- = \sin(\alpha-\beta)  \qquad  s_+ = \sin(\alpha+\beta)
\eeq{spmdef}
 and 
\beq
  {\bf D} = c_\alpha^2 c_\beta^2 G^1_{--}G^1_{-+} {G^2_{+-}\over G^2_{--}}
   + c_\alpha^2 s_\beta^2  G^1_{--} G^1_{++}
   + s_\alpha^2 c_\beta^2 G^1_{+-} G^1_{-+} 
   + s_\alpha^2 s_\beta^2 G^1_{+-}G^1_{++} {G^2_{--}\over G^2_{+-}}  
\eeqn
is a positive definite factor. 
This potential switches  from repulsive to attractive according to
the sign of  $s_-s_+$.   In the repulsive  case, the minimum is at
$s_W = 0$, in the attractive case, the minimum is at $s_W = 1$, so
there is no interesting competition here  that allows Higgs vacuum
expectation value to be arbitrarily small. 

\subsection{Case III}

For case III, we begin with the IR boundary conditions in
\leqn{casethree} and add UV boundary conditions, covering the same 16
possibilities as in the previous section.

As in the previous section, the first four cases, with equal boundary
conditions in the UV for both fermion multiplets, have potentials
independent of $\alpha$ and $\beta$.   The cases with all $+$ and all
$-$ boundary conditions in the UV give potentials equal to zero.  The
case
\beq
    \pmatrix{+ & +\cr
                          - &   - \cr} \qquad
\pmatrix{ + & -\cr   - &    + \cr} 
\eeq{firstinthree}
gives
\beq
    V(s_W,c_1, c_2) =  V_A(s_W,c_1) + V_R(s_W,c_2) \ , 
\eeqn
precisely the case with competition analyzed in Section 6.1.   The
last case 
\beq
    \pmatrix{- & +\cr
                         + &   - \cr} \qquad
\pmatrix{ - & -\cr  + &    + \cr} 
\eeq{secondinthree}
 gives a similar result.

The next set of cases have a potential that depends on one but not
both mixing angles.   The first example is
\beq
    \pmatrix{+ & +\cr
                          + &   - \cr} \qquad
\pmatrix{ - & -\cr   + &    + \cr}  \ .
\eeq{thirdinthree}
It is straightforward to work out the potential using the methods
already described.  We have 
\beq
\CC =  \pmatrix{  c_\alpha c_W G^1_{--}& - c_\alpha s_W G^1_{-+} & -
  s_\alpha c_W G^2_{-+}& s_\alpha s_W G^2_{--}    \cr
  s_W G^1_{--} & c_W G^1_{-+} & 0 & 0
 \cr  s_\alpha c_W G^1_{+-} & - s_\alpha s_W G^1_{++} & c_\alpha c_W G^2_{++} &
 -c_\alpha s_W G^2_{+-}   \cr
 0 & 0 & s_W G^2_{-+} & c_W  G^2_{--} \cr}\ .
\eeqn
Computing the determinant and assembling the Coleman-Weinberg
potential, we find
\beq 
 V = - 2 \int {d^4p\over (2\pi)^4}  \log \biggl[ 1 +  {s_W^2\over p^2
   z_0 z_R } { c_\alpha^2 G^1_{--}G^1_{-+}  - s_\alpha^2 G^2_{--}
   G^2_{-+}\over {\bf D}}  \biggr]\ ,
\eeq{Vforcasethree}
where now
\beq
 {\bf D} = c_\alpha^2 G^1_{--}G^1_{-+}  G^2_{--}G^2_{++} +  s_\alpha^2 G^2_{--}
   G^2_{-+}  G^1_{-+}G^1_{+-}  \ . 
\eeqn
This potential interpolates between the attractive 
case, for $s_\alpha
= 0$, and the repulsive case, for $s_\alpha = 1$.  However, for almost
all values of $s_\alpha$, the potential is monotonic and so is minimized
at $s_W = 0$, for
smaller values of $s_\alpha$ or at $s_W = 1$, for larger values of $s_\alpha$.
To understand this better, examine the
first two derivatives of \leqn{Vforcasethree}. These are 
\beqa
{\del V\over \del( s_W^2)} \biggr|_0  &=& -2 \int {d^4p\over (2\pi)^4}\ \biggl[  {1\over p^2
   z_0 z_R } { c_\alpha^2 G^1_{--}G^1_{-+}  - s_\alpha^2 G^2_{--}
   G^2_{-+}\over {\bf D}} \biggr]  \CR
{\del^2 V\over \del ( s_W^2)^2} \biggr|_0  &=& +2 \int {d^4p\over (2\pi)^4}\ \biggl[  {1\over p^2
   z_0 z_R } { c_\alpha^2 G^1_{--}G^1_{-+}  - s_\alpha^2 G^2_{--}
   G^2_{-+}\over {\bf D}} \biggr]^2   \ . 
\eeqan
The $s_W^4$ is always positive, but, when the $s_W^2$ term vanishes,
the $s_W^4$ 
term has almost the same zero  and is doubly suppressed.
Thus, this case has a second order phase transition where the vacuum
expectation value of the Higgs field goes to zero, but it occurs only
in an extremely fine-tuned interval of  $s_\alpha$. 

The other three cases in which $V$ depends on one mixing angle are
related to this case by exchanging $+ \leftrightarrow -$ boundary
conditions and exchanging the two fermion multiplets, by
interchanging top and bottom within each representation and sending
$\alpha \to \beta + \pi/2$, or by both of these operations.   All four
cases then have the behavior just described.

The remaining cases with this choice of IR boundary condition can all
be described as cases of 
\beq
    \pmatrix{+ & +\cr
                          + &   - \cr} \qquad
\pmatrix{ - & -\cr   - &    + \cr} 
\eeq{lastinthree}
with arbitrary values of the mixing angles $\alpha$ and $\beta$.  We
can get a feeling for the result by considering the special cases: (a)
For $\alpha = \beta = 0$, the value of the Coleman-Weinberg potential
is zero; (b) if $\alpha = \pi/2$, $\beta = 0$,  then both fermions are
in the repulsive case, (c) if $\alpha = 0$, $\beta = \pi/2$, both
fermions are in the attractive case.   Then there will be no
competition between the two representations, but the minimum of the
potential will swing back and forth between $s_W = 0$ and $s_W = 1$
according to the values of $\alpha$ and $\beta$.

The precise form of the potential can be worked out as in the previous
cases.  The result is
\beq 
 V = - 2 \int {d^4p\over (2\pi)^4}  \log \biggl[ 1  - {s_W^2 \over p^2 z_0 z_R} 
{s_- s_+ (G^1_{--} G^1_{++} + G^2_{-+} G^2_{+-})\over {\bf D} } 
-  {s^2_Wc_W^2 \over (p^2 z_0 z_R)^2 } {s_-^2 \over {\bf D}} \biggr\} \biggr] \ ,
 \eeq{Vforcasethreelast}
with $s_-$, $s_+$ as in \leqn{spmdef} and where now
\beqa
{\bf D} & =&   c_\alpha^2 c_\beta^2 G^1_{--}G^1_{-+} G^2_{+-} G^2_{++}
+ c_\alpha^2 s_\beta^2  G^1_{--}G^1_{++} G^2_{--} G^2_{++}\CR
& & +  s_\alpha^2 c_\beta^2 G^1_{-+}G^1_{+-} G^2_{-+} G^2_{+-}
+ s_\alpha^2 s_\beta^2  G^1_{+-}G^1_{++} G^2_{--} G^2_{-+} \ .
\eeqa{caseIIIhard}
Note that, using \leqn{Wronskiantwo},
\beq
     G^1_{--} G^1_{++} + G^2_{-+} G^2_{+-} = G^1_{-+} G^1_{+-} +
     G^2_{--} G^2_{++} \ ,
\eeqn
so \leqn{Vforcasethreelast}  has the required symmetry between $\psi_1$ and
$\psi_2$. 
This potential has just the form described in the previous paragraph,
with zeros along lines where $s_-  = \sin(\alpha-\beta) = 0$.

\subsection{Case IV}

For case IV, we begin with the IR boundary conditions in
\leqn{casefour} and add UV boundary conditions, covering the same 16
possibilities as in the previous section.

The cases with both UV boundary conditions equal is again trivial,
giving potentials equal to $0$, $0$, $V_A(s_W,c_1)+V_A(s_W,c_2)$, and 
$V_R(s_W,c_1)+V_R(s_W,c_2)$ in the four cases.

The first case with one mixing angle is 
\beq
    \pmatrix{+ & + \cr + & - \cr} \qquad
	\pmatrix{- & + \cr +& - \cr}  \ .
\eeq{caseIVa}
For $\alpha = 0$, we have zero for the potential from $\psi_1$ and the
repulsive case $V_R(s_W,c_2)$ from $\psi_2$.   For $\alpha =
\pi/2$, the potential from $\psi_1$ is in the repulsive case
$V_R(s_W,c_1)$ and the potential from $\psi_2$ is zero.  This suggests
that the potential is always repulsive, with its minimum at $s_W =
0$. The precise form of the potential is
\beq 
 V = - 2 \int {d^4p\over (2\pi)^4}  \log \biggl[ 1 - {s_W^2 \over p^2 z_0 z_R} 
{c_\alpha^2 G^1_{--} G^1_{-+} + s_\alpha^2 G^2_{--} G^2_{-+}\over {\bf D} } 
\biggr] \ ,
 \eeq{VforcaseIVa}
with 
\beq
{\bf D} =  c_\alpha^2 G^1_{--}G^1_{-+} G^2_{-+} G^2_{+-}
+s_\alpha^2   G^1_{-+}G^1_{+-} G^2_{--} G^2_{-+} \ . 
\eeqn
This expression is clearly positive definite, with a zero at $s_W =
0$.   The second case with one mixing angle
\beq
    \pmatrix{+ & + \cr - & - \cr} \qquad
	\pmatrix{- & + \cr - & - \cr} 
\eeq{caseIVb}
is similarly always in the attractive case, with its minimum at $s_W =
1$.   The full expression for the potential is 
\beq 
 V = - 2 \int {d^4p\over (2\pi)^4}  \log \biggl[ 1 + {s_W^2 \over p^2 z_0 z_R} 
{c_\alpha^2 G^2_{+-} G^2_{++} + s_\alpha^2 G^1_{+-} G^1_{++} \over {\bf D} } 
\biggr] \ ,
 \eeq{VforcaseIVb}
with 
\beq
{\bf D} =  c_\alpha^2 G^1_{--}G^1_{++} G^2_{+-} G^2_{++}
+s_\alpha^2   G^1_{+-}G^1_{++} G^2_{--} G^2_{++} \ . 
\eeqn
The remaining two cases are related to these by reversing the top and
bottom rows.

The final case, with dependence on two mixing angles, is 
\beq
    \pmatrix{+ & + \cr + & - \cr} \qquad
	\pmatrix{- & + \cr - & - \cr}  \ .
\eeq{infourth}
The Coleman-Weinberg potential can be worked out as above; the result
is 
\beq 
 V = - 2 \int {d^4p\over (2\pi)^4}  \log \biggl[ 1  + {s_W^2 \over p^2 z_0 z_R} 
{s_- s_+ (G^1_{++} G^1_{--} - G^2_{++} G^2_{--})\over {\bf D} } 
+  {s^2_Wc_W^2 \over (p^2 z_0 z_R)^2 } {s_-^2 \over {\bf D}} \biggr\} \biggr] \ ,
 \eeq{Vforcasefour}
with $s_-$, $s_+$ as in \leqn{spmdef} and where now
\beqa
{\bf D} & =&   c_\alpha^2 c_\beta^2 G^1_{--}G^1_{-+} G^2_{+-} G^2_{++}
+ c_\alpha^2 s_\beta^2  G^1_{--}G^1_{++} G^2_{+-} G^2_{-+}\CR
& & +  s_\alpha^2 c_\beta^2 G^1_{+-}G^1_{-+} G^2_{--} G^2_{++}
+ s_\alpha^2 s_\beta^2  G^1_{+-}G^1_{++} G^2_{--} G^2_{-+} \ .
\eeqan
The form of this expression shows explicit competition between
$\psi_1$ and $\psi_2$.   Most of this can be understood by considering
limit points where the two fermions decouple from one another:   at
$\alpha = 0$, $\beta = 0$, both fermions have potential equal to zero;
at $\alpha = \pi/2$, $\beta = 0$, $\psi_1$ is in the repulsive case
while $\psi_2$ is in the attractive case; at $\alpha = 0$, $\beta =
\pi/2$, $\psi_1$ is in the attractive case
while $\psi_2$ is in the repulsive  case.

\begin{figure}
\centering
\includegraphics[width=4in]{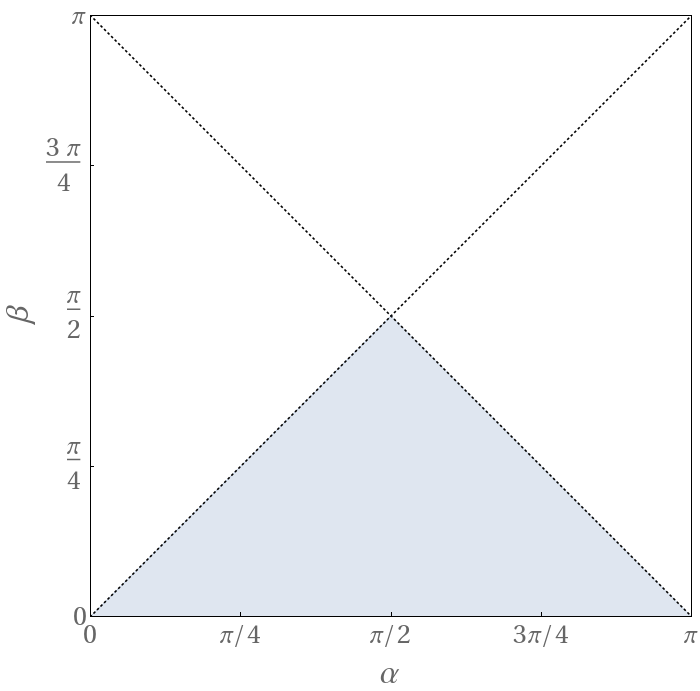}
\caption{Fundamental region of the $(\alpha,\beta)$ plane useful for  describing
  the phase diagram of the model \leqn{infourth}.}
\label{fig:IVtriangle}
\end{figure}

To understand the full dynamics of this model, it is useful to reduce
it to the minimal region of the $(\alpha,\beta)$ plane.    The
potential \leqn{Vforcasefour} depends only on $s_-s_+$ and $s_-^2$.
Then the potential takes the same value under the translations 
\beq
     \alpha \to \alpha + \pi,\   \beta \to \beta   \quad \mbox{and} 
 \quad  \alpha \to \alpha,\   \beta \to \beta +\pi \ . 
\eeqn
and under the reflection
\beq 
   \alpha\to - \alpha, \ \beta \to -\beta
\eeqn
This implies that the fundamental region for $(\alpha,\beta)$ is the
triangle
\beq
     0 < \alpha < \pi , \ 0 < \beta < \pi,    \   \alpha + \beta < \pi
     \ . 
\eeqn
Further, reflection across the line   $\alpha - \beta = 0$, that is,
\beq
     \alpha \leftrightarrow \beta \ , 
\eeqn
changes the sign of the competition term in \leqn{Vforcasefour}  and
so is equivalent to interchanging $\psi_1$ and $\psi_2$.   The full
dynamics of the model is then exhibited in the triangle shown in
Fig.~\ref{fig:IVtriangle}, with $\psi_1$ always in the repulsive case
and $\psi_2$ always in the attractive case.

The phase diagram shown in Fig.~\ref{fig:Aline} changes smoothly with
$\alpha$ and $\beta$ across this diagram.  Note that, while the
coefficient of $s_W^2$ in $V(s_W)$ can have either sign,
the coefficient of $s_W^4$ is always positive.   Then we will find a
line of second-order phase transitions where  $\del V/\del (s_W^2)$ is
zero.   At the bottom center of the triangle, $\alpha = \pi/2$, $\beta
= 0$, we have a case equivalent to that of Section 6.1.   There is a
curve of second-order phase transitions with its tip at $c_1 = 0 $, 
$c_2 = 0.2997$ (for $z_0/z_R=0.01$).
Across the bottom of the triangle, the critical value of $c_2$ for
$c_1 = 0$ increases slowly  from 0.2697 at $\alpha =
0$ to 0.2997 at $\alpha= \pi/2$ and back   to 0.2697   at  $\alpha = \pi$.  

The other two edges of the triangle have simple forms for $V(s_W)$.
Along the line $\alpha = \beta$, $V(s_W) = 0$.   Along the line
$\alpha + \beta = \pi$, $s_+ = 0$ and so the potential takes the simple
attractive form
\beq 
 V = - 2 \int {d^4p\over (2\pi)^4}  \log \biggl[ 1   
+  {s^2_Wc_W^2 \over (p^2 z_0 z_R)^2 } {s_-^2 \over {\bf D}}  \biggr] \ ,
 \eeq{VforcaseIVsimple}
with minima at $s_W^2 = c_W^2 = \half$.    In accordance with this,
the critical value of $c_2$ at $c_1 = 0$  varies along each 
horizontal  line with fixed  $\beta > 0$, tending to 0 as the left-hand
boundary is approached and to $\infty$ as the right-hand boundary is
approached.
The critical value at $\alpha = \pi/2 $ remains close to 0.3 for all
values of $\beta$.

\section{An example with relaxed fine-tuning}

We have now seen that the examples of the previous section can all be
understood in terms of the competition of fermion multiplets with
attractive and repulsive 
boundary conditions.   However, the only cases with a large $v/f$ hierarchy
were those in which the values of the parameters $c_1$ and $c_2$ were
adjusted to be close to a line of second-order phase transitions. In 
other words, the Coleman-Weinberg potential that we have encountered 
so far is always \textit{strongly} attractive or repulsive.  In most
of the parameter space, the value of 
of $v/f$ was
 not  affected by the competition, 
 and 
the potential was minimized at $s_W = 0$ or at $s_W = 1$. 

More quantitatively, among the terms in the potential expansion
\leqn{Vexpand}, the quadratic term $s_W^2$ almost always dominates
over the quartic term $s_W^4$ and therefore the overall sign of
$s_W^2$ simply determines the vacuum. For a non-trivial minimum, the
parameters $c_1$ and $c_2$ should be fine-tuned so that the overall
strength of $s_W^2$ becomes smaller than that of $s_W^4$. This implies
that for a natural explanation of a large $v/f$ hierarchy, a
\textit{weakly} repulsive fermion is 
required, which contributes to the Higgs 
potential only at the quartic level without the quadratic term.

\begin{figure}
\centering
\includegraphics[width=5in]{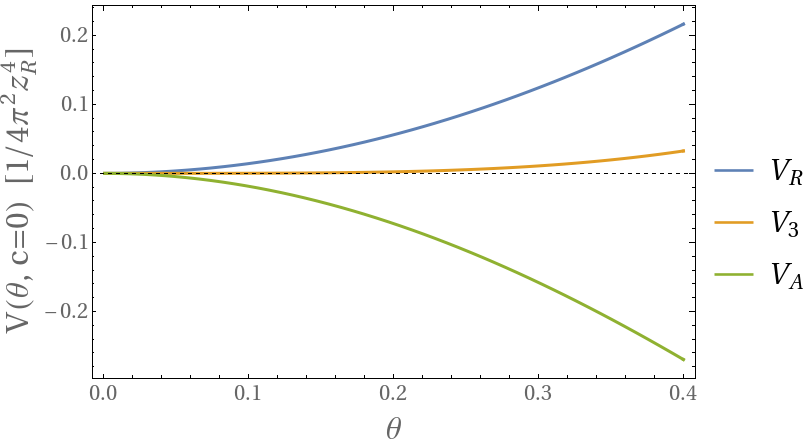}
\caption{The shape of $V_A, V_R, V_3$ near $s_W=0$. All three potentials are for $c=0$.}
\label{fig:nearzero}
\end{figure}

Here is an example:  Consider a fermion multiplet in the triplet
representation of $SU(2)$ 
with boundary conditions
\beq
	\psi_3 \sim \pmatrix{q_{+} \cr q_0 \cr q_{-} \cr} \sim \pmatrix{+ & - \cr - & - \cr - & + \cr}  \ ,
\eeq{tripletbc}
where $(q_+, q_0, q_-)$ are eigenstates of the generator $t^3$. If the
Goldstone boson $\VEV{A_5^2 }$ connected the two fermions $q_+$ and
$q_-$, 
the triplet would generate a repulsive potential. However, the form of
the generator is
\beq
	t^2 = {1 \over \sqrt{2}} \pmatrix{0 & -i & 0 \cr i & 0 & -i \cr 0 & i & 0 \cr}  \ , 
\eeq{triplett2}
and it only connects $q_+ \leftrightarrow q_0$ and $q_0
\leftrightarrow q_-$. Then the Coleman-Weinberg potential must be flat
in the $t^2$ direction, at least in the leading order. The same
applies to $\VEV{A_5^1}$. Indeed, the matrix $U_W$ acting on  $\psi_3$ has
the 
form
\beq
U_W = \pmatrix{ c^2_W  & -s_{2W}/ \sqrt{2} & s^2_W  \cr s_{2W}/
  \sqrt{2} & c_{2W} &  -s_{2W}/ \sqrt{2} \cr s^2_W  & s_{2W}/ \sqrt{2}
  &  c^2_W \cr } \ , 
\eeq{UWtriplet}
where  $s_{2W} = \sin 2\theta_W$.    The Coleman-Weinberg potential
from this multiplet is 
\beqa
    V_3 (s_W,c) =  -2 \int{d^4 p\over (2\pi)^4}  \log \biggl[ 1 - { s_W^4
    \over   p^2 z_0 z_R    G_{-+}  G_{+-} } \biggr] \ . 
\eeqa{tripletV}
This potential has no $s_W^2$ term and is repulsive in quartic
order. Fig.~\ref{fig:nearzero} shows the shape of the three potentials
$V_A, V_R, V_3$ near $s_W=0$, all for $c = 0$. We can see $V_3$ is indeed only \textit{weakly} repulsive.

\begin{figure}
\centering
\includegraphics[width=4in]{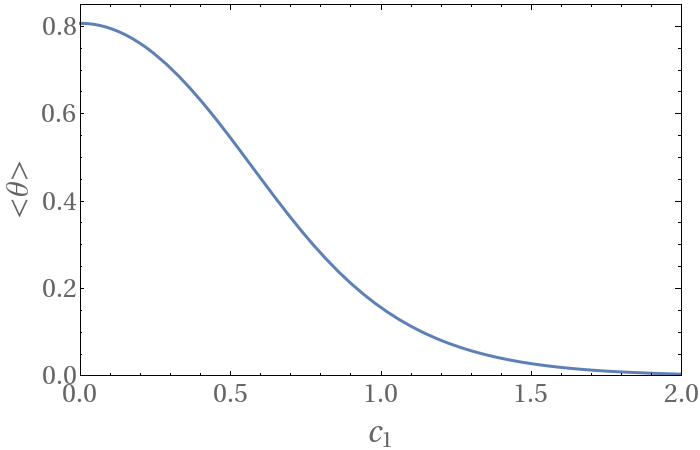}
\caption{The minimum of the potential $V_A(s_W, c_1) + V_3(s_W,
  c_3=0)$, shown as a function of $c_1$.}
\label{fig:withtriplet}
\end{figure}

To study the effect of the new triplet $\psi_3$ on the phase diagram,
first consider a system with two fermion multiplets, an attractive
doublet $\psi_A$ and the triplet $\psi_3$. Fig.~\ref{fig:withtriplet}
shows the minimum $\VEV{\theta}$ of the Coleman-Weinberg potential as
a function of $c_1$ with $c_3=0$ fixed.  This theory is always in the
broken phase $s_W > 0$, as it must be, but for $c_1 > 0.3$ the large contribution 
to the quartic term from $\psi_3$ multiplet pushes the minimum of the potential to small values.

\begin{figure}
\centering
\includegraphics[width=5in]{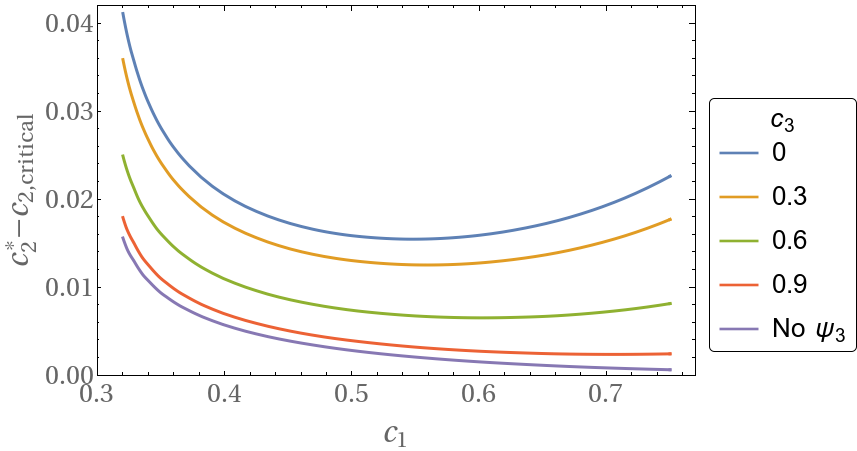}
\caption{Separation of $c_2^*$ from its critical value  $c_{2,critical}$ on the line of 
phase transition corresponding to   $\VEV{\theta}=0.1$.
These values are shown as a function of $c_1$, where the lowest curve
has the $\psi_3$ omitted (or $c_3\to \infty$) and the highest curve has
$c_3 = 0$.}
\label{fig:threefermion}
\end{figure}

More generally, we can use the multiplet $\psi_3$ to lower the degree
of fine-tuning needed to achieve a small value of $v/f$ in a system
with competition between attractive and repulsive fermion multiplets.
Consider  a model with $\psi_A$ and $\psi_R$ fermions as in Section
6.1, and add the multiplet $\psi_3$.
The position of the line of phase transitions does not change, since
$\psi_3$ contributes only quartic terms, but the presence of the
quartic term from $\psi_3$ can expand the region where $v/f$ is small.
 In Fig.~\ref{fig:threefermion},  we vary the parameter $c_3$ from
 high values to $c_3 = 0$ and  show the values of $c_1$ and
 $c_2$ for which $\VEV{s_W^2}=0.01$,  a value sought in realistic RS
 models.   The vertical axis is a measure of the fine-tuning
 needed to achieve $v/f \ll 1$.

It is interesting that  the  multiplet $\psi_3$ includes a right-handed zero mode. By
coupling it weakly to other fermions through boundary conditions at
$z_0$, we can give this fermion a small mass without disrupting the
Coleman-Weinberg potential. An interesting possibility for a realistic
model is then to introduce right-handed quarks and leptons in the
weakly repulsive multiplets and connect them at the UV boundary to the
left-handed doublets. This will generate fermion masses much smaller
than the top quark mass while simultaneously making  a 
$v/f$ hierarchy  more natural.

\section{Conclusions}

In this paper, we reviewed the formalism for fermions and gauge
fields in the RS geometry and the potential for fermion condensation.
We presented a simple formula, implementing ideas of Falkowski, for
computing the Coleman-Weinberg potential for the Higgs field.  Using
this formula, we explored the idea of competition between fermion
multiplets with different boundary conditions and presented strategies
for achieving the hierarchy  $v/f \ll 1$ needed in realistic models.

We hope that these tools will be useful for the construction of
realistic RS models with bulk fermions and gauge fields which could
provide predictive models of strongly coupled Higgs bosons.  In a
forthcoming paper, we will apply the methods discussed here to an
illustrative models of electroweak symmetry breaking driven by top
quark condensation~\cite{YPtwo}.

\appendix

\section{Basic formalism for fermions in RS}

In this appendix, we present details of our formalism for fermion
fields in RS.    We begin in Minkowski space.  Capital letters denote
5-dimensional indices, taking the values 0,1,2,3,5, with $M, N,
\ldots$ for world indices and $A, B, \ldots$, for tangent-space indices.  Lower-case
letters denote 4-dimensional indices. We use the metric
\leqn{RSmetric}. After deriving  the equations of motion, and after
gauge fixing in the case of vector bosons, 
we go to Euclidean space by the continuation $p^0\to i
p^0$,   $p^2\to
-p^2$. 

The Dirac action in RS is
\beq
S = \int d^4 x dz  \biggl(\sqrt{-g}\  \bar \Psi  \bigl[ i e^M_A \gamma^A {\cal
  D}_M - m \bigr] \Psi   - \bar {\cal K} \Psi - \bar \Psi {\cal K} \biggr)\ , 
\eeqn
where ${\cal D}_M$ is the gravity- and gauge-covariant derivative and
$e^M_A =  kz \delta^M_A$ for the metric \leqn{RSmetric}.  We 
denote  the gauge-covariant derivative as $D_M$; then ${\cal D}_M =
D_M + \half \omega_M{}^{AB}\Sigma_{AB}$.  The 
nonzero terms in the spin
connection are
\beq 
     \omega_m{}^{a5} = - \omega_m{}^{5a} =  {1\over z} \delta^a_m \ . 
\eeqn
We divide the 4-component Diract field $\Psi$ into two
2-component fields as in \leqn{fermiondec}, 
\beq
    \Psi = \pmatrix{ \psi_L \cr \psi_R \cr }   \  ,
\eeqn
using the basis of Dirac matrices
\beq
    \gamma^a =  \pmatrix{ & \sigma^a \cr \bar \sigma^a &  \cr } \ \
    \mbox{and} \  \ 
   \gamma^5 = - i \Gamma  \equiv - i \pmatrix{ -1 &  \cr & 1 \cr}   \ .
\eeqn
The matrix $\Gamma$ denotes the   4-dimensional chirality.  With these
conventions,
the Dirac action takes the form
\beqa
S &=& \int d^4 x dz \biggl( {1\over (kz)^4} \biggl[ \psi^\dagger_L  i
\bar\sigma^m D_m \psi_L  +\psi^\dagger_R i \sigma^m D_m \psi_R \CR 
& & \hskip  0.2in + \psi^\dagger_L \bigl( D_5 - {2\over z} -
{c\over z} \bigr) \psi_R  + \psi^\dagger_R \bigl(- D_5 +  {2\over z} -
{c\over z} \bigr) \psi_L  \biggr]  - \bar {\cal K} \Psi - \bar \Psi {\cal K} \biggr)\ .
\eeqan
Let 
\beq
   {\bf D} =  D_5 - {2+c \over z}   \ ,   \qquad   \bar {\bf D}=  D_5 - {2-c
     \over z}  \  .
\eeqn
Then the homogeneous equations of motion for $\Psi$ are
\beq 
	{1 \over (kz)^4}\pmatrix{ -\bar  {\bf D} & i \sigma^m D_m \cr i \bar \sigma^m D_m & {\bf D} \cr } 
	\pmatrix{ \psi_L \cr \psi_R  \cr } = 0   \ .
\eeq{psiLRfirst}

In gauge-Higgs unification,  we assume that the background gauge field has the form
 \beq
    A_M^a = (0,0,0,0, A_5^a(z) )  
\eeq{GHback}
 In this case, we can Fourier analyze in the 4 extended dimensions, so
 that  $i\bar \sigma^m D_m \to \bar \sigma \cdot p$,  $i\sigma^m D_m
 \to  \sigma \cdot p$.  Then we see that these fields obey 
 \beq 
	\Delta_{\Psi}(p^2) \Psi \equiv {1 \over (kz)^4}\pmatrix{ -\bar  {\bf D} & \sigma \cdot p \cr \bar \sigma \cdot p & {\bf D} \cr } 
	\pmatrix{ \psi_L \cr \psi_R  \cr } = 0
\eeq{psiDelta1}
The contribution of a fermion to the Coleman-Weinberg potential is then
\beq
    V_{\Psi} = - \int{ d^4 p \over (2\pi)^4} \log \det
    \Delta_{\Psi}(-p^2)  \ . 
\eeq{finalVforpsi}
This is the precise expression for the $\det(\Psi)$ term in \leqn{CWpot}.

We can eliminate either $\psi_L$ or $\psi_R$ from \leqn{psiDelta1}.
Once this is done, the remaining field obeys 
\beqa
   \Delta_L(p^2) \psi_L \equiv  \bigl( p^2 + {\bf D}\bar{\bf D}
   \bigr) \psi_L &=& 0 \CR
  \Delta_R(p^2) \psi_R \equiv  \bigl( p^2 +\bar  {\bf D} {\bf D}
   \bigr) \psi_R &=& 0 
\eeqa{psiDelta2}
Up to possible contributions from zero modes, $\Delta_L(p^2)$ and
$\Delta_R(p^2)$ have the same spectrum.   The operators $\Delta_L(p^2)$ and
$\Delta_R(p^2)$ include no spin matrices and can be thought of as
applied to single-component fields.   Then we can rewrite the
Coleman-Weinberg potential as 
\beq
    V_{\Psi} = - 2  \int{ d^4 p \over (2\pi)^4} \log \det
    \Delta_{L}(-p^2)  = - 2  \int{ d^4 p \over (2\pi)^4} \log \det
    \Delta_{R}(-p^2)\ .
\eeq{finalVforpsifinal}
The factor 2 counts the 2 spin degrees of freedom. 
This is the more precise expression for the $\det(\Psi)$ term in \leqn{CWpot}.

For $A_5(z) = 0$, the homogeneous equations \leqn{psiDelta2}
are solved by 
\beq
 \psi_L \sim   z^{5/2} ( J_{c+1/2}(pz), Y_{c+1/2}(pz) )   \qquad   \psi_R\sim  z^{5/2}
    ( J_{c-1/2}(pz), Y_{c-1/2}(pz) )   \ . 
\eeq{psieqnrel}
Standard identities for Bessel functions imply that these solutions
are interchanged by ${\bf D}$ and $\bar{\bf D}$, when $A_5(z) = 0$.  For example,
\beqa
  \bar  {\bf D}  (z^{5/2} J_{c+1/2}(pz)) & = &p \, (z^{5/2} J_{c-1/2}(pz)) \CR
   {\bf D}  (z^{5/2} J_{c-1/2}(pz)) & = &- p\, (z^{5/2} J_{c+1/2}(pz)) \ .
\eeqa{besselRules}

To calculate the Coleman-Weinberg potential, we must continue these
equations to Euclidean space.  The continuation of \leqn{psiDelta1} is 
 \beq 
	\Delta_{\Psi}  \Psi \equiv {1 \over (kz)^4}\pmatrix{ -\bar
          {\bf D} & i\sigma \cdot p \cr i \bar \sigma \cdot p & {\bf D} \cr } 
	\pmatrix{ \psi_L \cr \psi_R  \cr } = 0
\eeq{psiDeltaE}
where now $\sigma\cdot p = p^0 + i \vec\sigma\cdot \vec p$,
$\bar\sigma\cdot p = p^0 - i \vec\sigma\cdot \vec p$.
The operators ${\bf D}$, $\bar{\bf D}$ have the action on the $G$
functions \leqn{bigGdef}
\beq
    \bar{\bf D}_z \, z^{5/2}G_{+\beta}(z,z') = p  z^{5/2}
    G_{-\beta}(z,z') \qquad     {\bf D}_z \, z^{5/2}G_{-\beta}(z,z') = -p  z^{5/2}
    G_{+\beta}(z,z') \ .
\eeq{besselmorerules}

The field $\Psi$ has four Green's functions, 
\beqa 
   \G_{LL} (z,z',p) = \VEV{\psi_L(p,z) \psi_R^\dagger(-p,z')}  &\qquad&
   \G_{LR}(z,z',p)  = \VEV{\psi_L(p,z) \psi_L^\dagger(-p,z')} \CR
  \G_{RL} (z,z',p) = \VEV{\psi_R(p,z) \psi_R^\dagger(-p,z')}  &\qquad&
   \G_{RR} (z,z',p) = \VEV{\psi_R(p,z) \psi_L^\dagger(-p,z')} 
\eeqan
which are interconnected through the equations
\beq 
	\Delta_{\Psi} \G(z,z',p) =   \delta(z-z') {\bf 1}
\eeq{greenpsi}
and similar equations with operators applied to the right and acting
on $z'$. 
If the fermion field has multiple gauge components   $\Psi^A$, these equations become matrix
equations. For example, ${\cal G}^{AB}_{LL}$ will have the form
\beqa
   {\cal G}_{LL}^{AB}(z,z',p)   &=& p^2 z_R k^4 (z z')^{5 \over 2}  [ {\bf A}^{AB} G^{(A)}_{+,-A_R}(z,z_R) G^{(B)}_{-,-B_R}(z',z_R) \CR
 & & \hskip -0.5in -  \cases{
         \delta^{AB} A_R G^{(A)}_{+,A_R}(z,z_R) G^{(A)}_{-,-A_R}(z',z_R)]  &    $ z < z'$ \cr
         \delta^{AB} A_R G^{(A)}_{+,-A_R}(z,z_R)G^{(A)}_{-,A_R}(z',z_R)] &    $ z > z'$ \cr }  \ .
\eeqa{generalGreen}
In this equation, $A_R$ represents the boundary condition of the field
$A$ at $z=z_R$. That is, $A_R=+$ if the field $A$ has $+$ boundary
condition on IR brane, and $A_R=-$ if otherwise. $A_R=\pm$ implies
$-A_R=\mp$. 
In the second line, $A_R$ denotes a factor $\pm1$ depending on the
sign of $A_R$.  
We can obtain ${\cal G}^{AB}_{LR}$, ${\cal G}^{AB}_{RL}$, and ${\cal
  G}^{AB}_{RR}$, 
using \leqn{greenpsi}, the similar equation acting on $z'$,  and
\leqn{besselmorerules}.
In particular, ${\cal
  G}^{AB}_{RR}(z,z') $ has the same form with the first indices of the
$G$ functions reversed $+\leftrightarrow -$ from \leqn{generalGreen}.  

Because $\psi_L$ and $\psi_R$ are interconnected, it is not consistent to place
separate boundary conditions on these fields.  Instead,  it is
sufficient to place the boundary conditions
\beq
     + \ :   \ \   \psi_R = 0   \quad \mbox{or} \quad    -  \ :\ \
     \psi_L =
     0 
\eeq{psibcs}
A zero mode in $\psi_L$ requires   $(++)$ boundary conditions;
a zero mode in $\psi_R$ requires  $(--)$ boundary conditions. 

  Note
that the equations for $L$ and $R$ are interchanged by the interchange
of boundary conditions $+\leftrightarrow -$ and the interchange ${\bf
  D} \leftrightarrow \bar {\bf D}$, or equivalently, $c
\leftrightarrow -c$.  After these two interchanges, the fermion field
will have the same functional determinant.

\section{Basic formalism for gauge fields in RS}

In this appendix, we present details of our formalism for gauge fields
in RS.   Conventions for the 5-dimensional space are as in Appendix A.

The gauge field action in RS  is 
\beq 
S = \int d^4 x dz\ \biggl(\sqrt{-g}  \biggl[ - {1\over 4} g^{MP} g^{NQ} F^a_{MN}
F^a_{PQ} \biggr] -   {\cal J}^M A_M  \biggr)   \ .
\eeq{gaugeaction}

In our formalism, the Higgs field is a background gauge field, so we
will quantize in the Feynman-Randall-Schwartz  background field gauge~\cite{RandallSchwartz}.
Expand
\beq    
   A_M^a \to  A_M^a(z)  + \A_M^a  \ , 
\eeqn
where, on the right,  $A_M^a$ is a fixed background field, 
\beq
    A_M^a = (0,0,0,0, A_5^a(z) )  
\eeqn
as in \leqn{GHback}, 
 and $\A_M^a$ is a fluctuating field.  Let $A_M = A_M^a t^a$ and $F_{MN} =
F^a_{MN} t^a$, where $t^a$ are the generators of the gauge group, and let  $D_M$ be the covariant
derivative containing the background field only.   Then the linearized
form for the field strength is
\beq
    F_{MN}  =   D_M \A_N -   D_N \A_M  \ .
\eeqn
  
After inserting the metric \leqn{RSmetric} and performing some integrations by parts, the linearized gauge action becomes
\beqa
   S &=&  \int d^4 x dz \  \biggl(    \half {1\over kz} \biggl[ \A^n
   D^m D_m \A_n -  \A_m D^n D^m \A_n -  \A^n kz D_5 {1\over kz} D_5
   A_n \CR
& & \hskip 0.4in  - \A_5 D^m D_m \A_5 + 2 \A_5 D^m D_5 \A_m \biggr]   -
\J^m\A_m + \J_5 \A_5 \biggr) \ . 
\eeqan
Here and in the following, raised and lowered indices are contracted
with the Lorentz metric $\eta^{mn}$. 
Following \cite{RandallSchwartz}, introduce the gauge-fixing term
\beq
    S_{GF} =   \int d^4 x dz {1\over (kz)^5} \biggl[ -{1\over 2\xi}
\biggl( (kz)^2 D^m \A_m - \xi (kz)^3 D_5 {1\over kz}
     \A_5\biggr)^2 \biggr] \ . 
\eeqn
Then
\beqa
 S + S_{GF} & = &  \int d^4 x dz \  \biggl(    \half {1\over kz}
 \biggl[ \A_m \bigl( \eta^{mn} D^2  - \eta^{mn} kz D_5 {1\over kz} D_5 -   D^m
 D^n (1 - {1\over \xi}) \bigr) \A_n \CR 
  & & \hskip 0.4in  +  \A_5 \bigl( - D^2 + \xi  D_5 kz D_5 {1\over kz}
  \bigr) \A_5 \biggr]   -
\J^m\A_m + \J_5 \A_5 \biggr) \ . 
\eeqan
The linearized ghost action is 
\beq
  S_{ghost} = \int d^4 x dz \  \biggl(  {1\over kz}
 \biggl[ \bar c \bigl( - D^2 + \xi  kz D_5 {1\over kz} D_5\bigr) c
 \biggr]  -  \bar {\cal C} c - \bar c {\cal C} \biggr)\ .
\eeqn
These formulae simplify for $\xi = 1$.  The homogeneous equations for 
the gauge field components are 
\beqa
   \Delta_G(p^2) \A_m(z,p) &\equiv& {1 \over kz} \bigl( p^2 + z D_5 {1\over z} D_5
   \bigr) \A_m(z,p) = 0 \CR
\Delta_5(p^2) \A_5(z,p) &\equiv& {1 \over kz} \bigl( p^2 + D_5 z D_5 {1\over z} \bigr) \A_5(z,p) = 0\CR
 \Delta_c(p^2)  c(z,p) &\equiv& {1 \over kz} \bigl( p^2 + z D_5 {1\over z} D_5
 \bigr) c(z,p) = 0 \ .
\eeqa{AAfiveceqs}

Up to possible contributions from zero modes, $\Delta_G(p^2)$, $\Delta_5(p^2)$ and
$\Delta_c(p^2)$ have the same spectrum for consistent boundary
conditions, as defined below.   Then when we integrate out the
fields $\A_m$, $A_5$, and $(c,\bar c)$  we find
\beq
    (\det \Delta_G)^{4/2} (\det \Delta_5)^{1/2} (\det \Delta_c)^{-1} =
    (\det\Delta_G)^{3/2} \ .
\eeq{prodofdets}
The contribution of a gauge boson to the Coleman-Weinberg potential is then
\beq
    V_{G} =   +{3\over 2}  \int{ d^4 p \over (2\pi)^4} \log \det
    \Delta_G(-p^2)  \ . 
\eeq{finalVforA}
This is the precise expression for the $\det(A)$ term in \leqn{CWpot}.
The operators $\Delta_G$, $\Delta_5$ are related to the operators
$\Delta_L$, $\Delta_R$ defined in \leqn{psiDelta2} for fermion fields
with $c = \half$, by
\beq
  \Delta_L =  z^{3/2} (kz \Delta_G) {1\over z^{3/2} }\qquad 
    \Delta_R = z^{3/2} (kz \Delta_5)  {1\over z^{3/2} }  \ . 
\eeq{gaugetofermion}
 Thus, the calculation of the determinant of $\Delta_G$ and $\Delta_5$ for gauge
fields are special cases of the determinant calculation for fermion fields.

For $A_5(z) = 0$, the homogeneous equations \leqn{AAfiveceqs}
are solved by 
\beq
 A_m \ , \ c \sim   z^{1} ( J_1(pz), Y_{1}(pz) )   \qquad   A_5 \sim  z^{1}
    ( J_{0}(pz), Y_{0}(pz) )   \ . 
\eeq{Aeqnrel}
Standard identities for Bessel functions imply that these solutions
are interchanged by the action of $\del_5$ and $kz \del_5 (1/kz)$.   

 The Green's functions for gauge fields are
\beqa
    \VEV{ \A_m(z,p) \A_n(z',-p)} &=&   \eta_{mn} \G(z,z',p) \CR
    \VEV{ \A_5(z,p) \A_5(z',-p)} &=&  \G_5(z,z',p) \CR
    \VEV{ c(z,p) \bar c(z',-p)} &=&  \G_c(z,z',p)  \ .
\eeqan
These 
satisfy the differential equations in $z$
\beqa
	\Delta_G(p^2)  \G(z,z',p) &=&  \delta(z - z') \CR
	\Delta_5(p^2)  \G_5(z,z',p) &=&  \delta(z - z') \CR
	\Delta_c(p^2)  \G_c(z,z',p) &=& \delta(z - z')  \ .
\eeqa{AGreens}

The solutions to the gauge field equations 
are interrelated by 
\beq
       \A_5(z) =    D_5 \A_m(z)  \qquad    \A_m(z)  =  kz D_5 {1\over
         kz}  \A_5  \ .
\eeqn
These transformations interchange the boundary conditions
\beqa
     + \ :   \ \    D_5 \A_m(z) = 0  &  \qquad &    -  \ :\ \  \A_m(z) =
     0 \CR
     - \ :   \ \      \A_5(z) = 0  &  \qquad &    + \ :\ \  kz D_5
     ({1\over kz} \A_5(z) )  =
     0 \CR
\eeqa{Abcs}
If $\A_m(z)$ is assigned the boundary condition $+$ (respectively,
$-$), then consistently $\A_5(z)$ must be assigned the boundary
condition $-$ (respectively, $+$).

\section{Proof of Falkowski's Theorem}

In this appendix, we provide a proof of Falkowski's Theorem
\leqn{FalkowskiThm} by
 explicit calculation of $A_5$ tadpole diagrams. In Appendices~A 
and B, we have obtained the operators 
$\Delta_{\Psi}$, $\Delta_G$, $\Delta_5$ and
$\Delta_c$ from the quadratic Lagrangian for fermion and 
gauge fields under the background field
\beq
    A_M^a = (0,0,0,0, A_5^a(z) ) \  .
\eeqn
Then, the effective potential from each field can be written as the
functional
 determinant of the corresponding  operator. 

In this derivation, we will turn on the gauge field $A_5^a$ along a
fixed direction in the adjoint representation.   Then we will simplify
by writing $A_5^a t^a = A_5 t$. 
 It will be important to remember that, while the mixing matrix
$U_M$ in \leqn{UUCformula} can mix fermion fields in a more arbitrary
way, the matrices $t$ and $U_W$ can only connect fermions in the same
gauge representations, which must therefore have the same value of
$c$.  (For gauge bosons, always $c = \half$.)    If $t$ is
proportional to the unit matrix, it generates a pure phase in $U_W$.
 This affects the determinant of $\CC$, but such a phase manifestly
 cancels out of the equation \leqn{secondGbc} and so has no effect on
 $\AA$.    In the proof of the theorem below, we can then assume that $t$
 generates no overall phase, 
\beq
    \sum_{AB} t_{AB} \delta^{AB} = 0 \ . 
\eeq{noUoneint}

Consider first the fermion case
\beq
    V_{\Psi} =   - \int{ d^4 p \over (2\pi)^4} \log \det \Delta_{\Psi} \ . 
\eeq{genericV}
where $p$ is the 4d Euclidean momentum. Varying $A_5(z)$, we find
\beqa
 \delta V_{\Psi}  &=&   
    - \int{ d^4 p \over (2\pi)^4}\  \tr \left[ \Delta_{\Psi}^{-1}
 {\delta \Delta_{\Psi} } \right]  \CR
    &=&  - \int{ d^4 p \over (2\pi)^4} \int_{z_0}^{z_R} dz  \
\tr \left[ \G(z,z)   {1\over (kz)^4}(-i g \Gamma \delta A_5(z)  t ) \right] \CR
    &=& - \int{ d^4 p \over (2\pi)^4} \int_{z_0}^{z_R} dz (-i g \delta
    A_5(z) ) 
    \ 
\tr \left[ {1\over (kz)^4} (-\G_{LL}(z,z)+\G_{RR}(z,z))  t \right] \
. \CR
\eeqa{arrangeV}

We can obtain the Green's
function by gauging $A_5$ away to the UV boundary, as explained in the
Section 3. The full Green's function is related to the Green's
function for $A_5(z) = 0$ by 
\beq
       \G(z,z') =     \exp[ -ig \int^{z_R}_z d\bar z  A_5(\bar z) t] \
 \G_0(z,z') \  \exp[ +ig \int^{z_R}_{z'} d\bar z  A_5(\bar z) t]  \ .
\eeqn
In our method of turning on $A_5$, this field is essentially Abelian,
and the exponential factors cancel out for $z = z'$.   Then we can use 
the expression of the Green's function from
\leqn{generalGreen} to evaluate \leqn{arrangeV}.  The trace part within the integrand 
becomes
\beqa
{\cal T} &=& {1 \over (kz)^4} \tr \left[ (-\G_{LL}(z,z)+\G_{RR}(z,z)) t \right] \CR
	&=&  2 p^2 z z_R \sum_{A,B} t_{BA} 
{\bf A}^{AB} \CR & &\hskip 0.2in  ( -G^{(A)}_{+,-A_R}(z,z_R) G^{(B)}_{-,-B_R}(z,z_R)
              + G^{(A)}_{-,-A_R}(z,z_R)  G^{(B)}_{+,-B_R}(z,z_R) )  \ .
\eeqa{Teval}
The factor 2 comes from the trace over spinor indices.  Note that the
terms 
with $\delta^{AB}$ in $\G^{AB}_{LL}(z,z)$ and $\G^{AB}_{RR}(z,z)$ are
identical in the symmetric limit $z\to z'$ and cancel each other.  Here and
in the rest of this appendix, summation over field indices $A$,
$B$. etc.  is
always explicitly shown. 

From \leqn{Ceval}, we have 
\beq
     \CC_{AC} =   (U_MU_W)_{AC}\,  G^{(C)}_{-A_0, -C_R}  \ .
\eeq{CevalA}   
Using  \leqn{secondGbc} and \leqn{generalGreen}, we can formally
solve for $\AA^{AB}$,
\beq
\AA^{AB} = \sum_{C,D}  ({\bf C}^{-1})_{AC} (U_MU_W)_{CD} D_R \delta^{DB} G^{(D)}_{-C_0, +B_R}
\ .
\eeqn
In last line of \leqn{Teval}, the states $A$ and $B$ are connected by
$t_{BA}$.  Then these states have the same value of $c$, and so we can
use \leqn{Wronskiantwo} to evaluate the  expression in parentheses,
\beqa
& & \hskip -0.8in  ( -G^{(A)}_{+,-A_R}(z,z_R) G^{(B)}_{-,-B_R}(z,z_R) + G^{(A)}_{-,-A_R}(z,z_R)
G^{(B)}_{+,-B_R}(z,z_R) )  \CR
&=& \cases{    0   &     $ A_R = B_R$ \cr
                 - 1/p^2z z_R & $ A_R = +$, $B_R = -$\cr
                + 1/p^2z z_R & $ A_R = -$, $B_R = +$\cr} \CR
&=&  ( B_R \delta_{A_R, -B_R}) / p^2 z z_R 
\eeqan
Assembling the pieces,
\beqa
{\cal T} &=& 2 \sum_{A,B,C} ({\bf C}^{-1})_{AC}
( (U_MU_W)_{CB} B_R
G^{(B)}_{-C_0,+B_R})\  
 t_{BA} \  (B_R \delta_{A_R, -B_R} )  \CR
 &=& 2 \sum_{A,B,C} \bigl(({\bf C}^{-1})_{AC} \ (U_MU_W)_{CB}  G^{(B)}_{-C_0,
-A_R}\bigr)\   t_{BA}  \  \delta_{A_R, -B_R}  \ .
\eeqa{Tevaltwo}
The two factors of $B_R$ cancel, and then we are very close to the
desired form.

There is one further issue:  The sum in \leqn{Teval} is taken only
over pairs $(A,B)$ such that $A_R = - B_R$.   We would like to extend
this to a sum over all pairs.  To do this, use the identity $\CC^{-1}
\CC = {\bf 1}$.  Writing this out using \leqn{CevalA}, we have
\beq
	\sum_C (\CC^{-1})_{AC} (U_M U_W)_{CB} G^{(B)}_{-C_0 ,-B_R}  =
        \delta_{AB}
\eeq{CCid}
For $(A,B)$ such that $A_R = B_R$, this has the same form as the
summand of \leqn{Tevaltwo}.    For $A\neq B$, \leqn{CCid} is zero and
we can add these terms to \leqn{Tevaltwo}.  For $A = B$, for which
necessarily $A_R = B_R$, the trace of this identity with $t_{AB}$  is
$\delta_{AB} t_{AB} = 0$, and so we can also add back those terms.  
 Then there is no change in \leqn{Tevaltwo}
if we extend the sum to all pairs $(A,B)$.

Finally, we find
\beqa
 \delta V_{\Psi}  &=&   
    -2 \int{ d^4 p \over (2\pi)^4}\ \sum_{A,B,C}  (\CC^{-1} )_{AC} (U_M
    U_W)_{CB} \left( -ig \int dz
    \delta A_5(z) t\right)_{BA} G^{(B)}_{-C_0,-A_R}\CR
&=&  
 -2 \int{ d^4 p \over (2\pi)^4}\ \sum_{A,B,C}   \CC^{-1}_{AC}  \
   \delta \CC_{CA}   \CR
&=&   
 -2 \int{ d^4 p \over (2\pi)^4}\  \delta \log \det \CC
\eeqan
Integrating up from $A_5(z) = 0 $,  we obtain the contribution of a fermion to the Coleman-Weinberg potential
\beq
    V_{\Psi} = -2 \int{ d^4 p \over (2\pi)^4} \log \det {\bf C}(p)
\eeqn
up to an additive, $A_5$-independent constant. This completes the
proof of 
Falkowski's Theorem for the fermion determinants.

In \leqn{gaugetofermion}, we pointed out that the  evaluations of the
gauge boson determinants were special cases of the evaluations of the
fermion determinants with $c= \half$.  Thus, this method of evaluation holds also for
the gauge boson determinants.

\section{Properties of the fermionic Coleman-Weinberg potentials for
   $SU(2)$ doublets}

In this appendix, we discuss the expansion of the canonical attractive
and repulsive potentials \leqn{VAdef} and \leqn{VRdef} for small
values of $s_W$. 

The symmetry under reversal of boundary conditions and
the sign of $c$ noted at the end of Appendix A implies that 
\beq
  V_A(s_W,-c) = V_A(s_W,c)  \qquad  V_R(s_W,-c) = V_R(s_W,c)  \ .
\eeqn
So, in this Appendix, we will restrict ourselves to $c \geq 0$. 

The repulsive case is more straightforward.   The integrand of
\leqn{VRdef} can be expanded under the integral
sign.   Then 
\beq
  V_R(s_W,c) =  {1 \over 4 \pi^2 z_R^4 } \left[ A_R(c)  s_W^2 + {1 \over 2} B_R(c) s_W^4 + \cdots \right] \ ,
\eeqn
as in \leqn{Vexpand}, where
\beqa
A_R(c) & = & \int_0^\infty dp\, p^3  \ { z_R^4 \over  p^2 z_0 z_R G_{-+}
  G_{+-}}  \CR
B_R(c)  & = & \int_0^\infty  dp\,  p^3\   { z_R^4 \over ( p^2 z_0 z_R G_{-+}
  G_{+-})^2}  \ . 
\eeqan
The functions $G_{-+}$, $G_{+-}$ increase exponentially with $p$
according to \leqn{Gasympt}, and so the integrals are convergent in
the UV.   In addition, these functions behave as $p \to 0$ as 
\beq 
    G_{-+} G_{+-}  =    {1\over p^2 z_0 z_R} (1  + {\cal O}(p)) \ , 
\eeqn
so the integrals are convergent in the IR. Also note that $A_R$ and $B_R$ depend only on the ratio $z_0/z_R$, not on $z_0$ or $z_R$ individually. There is a weak dependence on $z_0/z_R$ when  $z_0 \ll z_R$, the
case of interest to us.

For the representative case  $z_0/z_R = 0.01$, the
values of these coefficients at $c = 0 $ are 
\beq
  A_R(0) = 1.4078    \qquad     B_R(0) = 0.2169 \ , 
\eeqn
and the dependence on $c$ is qualitatively described by 
\beq
{ A_R(c)\over A_R(0)} \approx \exp[ - 2.9 c^2]  \qquad
{ B_R(c)\over B_R(0)} \approx \exp[ - 4.4 c^2]  \ .
\eeqn

For the attractive case, more care is necessary.   The functions
$G_{--}$, $G_{++}$  go to constants as $p\to 0$.  Let
\beq
   \GG =   z_0 z_R  G_{--}(0) G_{++}(0)  \ .
\eeqn
For $c = 0$ and $z_R \gg z_0$,  $\GG \approx z_R^2$.    The leading
coefficient in $V_A(s_W^2) $ is the convergent integral
\beq
  A_A(c) = \int_0^\infty dp\,  p^3\  { z_R^4  \over  p^2 z_0 z_R G_{--}
  G_{++}}  \ .
\eeqn
To evaluate the $s_W^4$ terms, differentiate $V_A$ twice with respect
to $s_W^2$, 
\beq
 {\del^2 V_A\over \del (s_W^2)^2 } =   \int dp\, p^3\  {z_R^4 
   \over  (p^2 z_0 z_R G_{--} G_{++} + s_W^2)^2}
\eeqn
and evaluate the integral by breaking it into two parts at a value
$\epsilon$ such that $ s_W^2 \ll \epsilon^2\GG  \ll  1$.  The
integral for $p < \epsilon$ can be evaluated directly.   The integral
for $p > \epsilon$ can be evaluated by adding and subtracting a term
that cancels the infrared divergence.  This gives
\beq
 {\del^2 V_A\over \del (s_W^2)^2 } =  {z_R^4 \over 2} \biggl[  {1\over
  ( \GG)^2 } (\log{1\over s_W^2} -  \gamma -1) + 
   \int_0^\infty {dp^2\over p^2} \ \bigl\{  { 1\over
   (z_0 z_R G_{--} G_{++})^2}  - {1\over (\GG)^2} e^{- \GG p^2} \bigr\}
   \biggr] \ . 
\eeqn
Integrating back, we find
\beq
  V_A(s_W,c) = {1 \over 4 \pi^2 z_R^4 } \left[ - A_A(c)  s_W^2 + {1 \over 2} B_A(c) s_W^4 + {1 \over 2} C_A(c)  s_W^4
  \log{1\over s_W^2} + \cdots \right] \ ,
\eeqn
as in \leqn{Vexpand}, where
\beqa
A_A(c) & = &  \int_0^\infty dp\,  p^3 \ { z_R^4 \over  p^2 z_0 z_R G_{--}
  G_{++}}  \CR
B_A(c)  & = &z_R^4 \biggl[ {1\over (\GG)^2} [ {1\over 4} -
  {\gamma\over 2} ] + \int_0^\infty {dp\over p} \ \bigl\{  {1\over
   (z_0 z_R G_{--} G_{++})^2} - {1\over (\GG)^2} e^{- \GG p^2} \bigr\}
   \biggr] \CR
C_A(c) &=&  {z_R^4 \over 2  (\GG)^2 }
\eeqan
These coefficients also have a weak dependence on $z_0/z_R$.

For the representative case  $z_0/z_R = 0.01$, the
values of these coefficients at $c = 0 $ are 
\beq
  A_A(0) = 1.8771    \quad     B_A(0) = 0.1958 \quad  C_A(0) =  0.5205  \ , 
\eeqn
and the dependence on $c$ is qualitatively described by 
\beq
{ A_A(c)\over A_A(0)} \approx \exp[ - 3.3 c^2]  \qquad
{C_A(c)\over C_A(0)} \approx \exp[ - 6.7 c^2]  \ ,
\eeqn
where $B_A(c)$ has a non-trivial dependence on $c$, with maximum at $B_A(0.1981)=0.2029$ and exponential suppression for large $c$. 

Again for $z_0/z_R = 0.01$, the solution to the equation
\beq
      A_A(c_1) =  A_R(0)  
\eeq{AAReq}
is 
\beq
     c_1 =   0.2997    \ .
\eeqn
This point gives the tip of the locus of second-order transitions in Fig.~\ref{fig:Aline}. Along the line of phase transitions, we can parametrize the total quartic term as a function of $c_1$. The coefficient of  $s_W^4 \log{1\over s_W^2}$ term is simply $C(c_1) = C_A(c_1)$. The coefficient of $s_W^4$ is well approximated by a linear equation,
\beq
B(c_1) = 0.41 - 0.99 (c_1-0.3) \quad \textrm{for} \quad 0.3 < c_1 < 0.6 \ ,
\eeqn
and approaches zero for large $c_1$.

\newpage

\Acknowledgements

We are grateful to Roberto Contino, Christophe Grojean, Yutaka
Hosotani, Yael Shadmi, and our colleagues in the SLAC Theory Group for discussions
of the ideas presented in this paper.
This work was supported by the U.S. Department 
of Energy under contract DE--AC02--76SF00515. JY is supported by a Kwanjeong Graduate Fellowship.

\end{document}